\documentclass[%
 reprint,
superscriptaddress,
 amsmath,amssymb,
 aps,
]{revtex4-2}

\usepackage{empheq}
\usepackage{hyperref}
\usepackage{graphicx}
\usepackage{dcolumn}
\usepackage{mathtools}
\usepackage{bm}

\usepackage[dvipsnames]{xcolor}
\usepackage{comment}
\usepackage{chemformula}

\newcommand\rev[1]{{\color{Black}#1}}

\begin{document}
\title{Runaway Transition in Irreversible Polymer Condensation with cyclisation}
\author{Maria Panoukidou}
\thanks{joint first author}
\affiliation{School of Physics and Astronomy, University of Edinburgh, Peter Guthrie Tait Road, Edinburgh, EH9 3FD, UK}
\author{Simon Weir}
\thanks{joint first author}
\affiliation{School of Physics and Astronomy, University of Edinburgh, Peter Guthrie Tait Road, Edinburgh, EH9 3FD, UK}
\author{Valerio Sorichetti}
\affiliation{Institute of Science and Technology Austria, 3400 Klosterneuburg, Austria}
\affiliation{Laboratoire de Physique Théorique et Modèles Statistiques (LPTMS), CNRS, Université Paris-Saclay, F-91405 Orsay, France}
\author{Yair Gutierrez Fosado}
\affiliation{School of Physics and Astronomy, University of Edinburgh, Peter Guthrie Tait Road, Edinburgh, EH9 3FD, UK}
\author{Martin Lenz}
\affiliation{Laboratoire de Physique Théorique et Modèles Statistiques (LPTMS), CNRS, Université Paris-Saclay, F-91405 Orsay, France}
\affiliation{PMMH, CNRS, ESPCI Paris, PSL University, Sorbonne Université,
Université de Paris, F-75005, Paris, France}
\author{Davide Michieletto}
\thanks{corresponding author, davide.michieletto@ed.ac.uk}
\affiliation{School of Physics and Astronomy, University of Edinburgh, Peter Guthrie Tait Road, Edinburgh, EH9 3FD, UK}
\affiliation{MRC Human Genetics Unit, Institute of Genetics and Cancer, University of Edinburgh, Edinburgh EH4 2XU, UK}

\begin{abstract}
The process of polymer condensation, i.e. the formation of bonds between reactive end-groups, is ubiquitous in both industry and biology. Here we study generic systems undergoing polymer condensation in competition with cyclisation. Using a generalised Smoluchowski theory, molecular dynamics simulations and experiments with DNA and ATP-consuming T4 ligase, we find that this system displays a transition, from a ring-dominated regime with finite-length chains at an infinite time to a linear-polymers-dominated one with chains that keep growing in time. Finally, we show that fluids prepared close to the transition may have widely different compositions and rheology at large condensation times.
\end{abstract}

\maketitle

\section{\label{sec:level1}Introduction}
Linear polymer condensation is the process by which two polymeric end groups react to form a bond. Beyond its relevance to industry~\cite{Rubinsteina}, \rev{and biotechnology~\cite{Oliynyk2022}}, it underpins the biophysics of DNA repair and cloning~\cite{Alberts2014}. \rev{In the absence of loop formation, polymer condensation will yield linear chains with average length $\langle l \rangle = 1/(1-p)$ where $p$ is the extent of the condensation reaction~\cite{Rubinsteina,Flory1936}}. However, looping, or cyclisation, is expected to be favourable in certain conditions~\cite{Cates2001,VafabakhshRezaHa2012,Zhou2012}. Several theories on reversible polymer condensation and experiments have, over the last decades, attempted to reach a consensus on whether the polymers in such systems will all eventually convert into rings or whether there always be a linear population at a large-time scale ~\cite{Jacobson1950,Flory1966,Suematsu1992,Dormidontova2004,Ercolani2008,Madeleine-Perdrillat2014a,DiStefano2019,Kricheldorf2020}. \rev{Despite this, the polymer physics and chemistry communities have not yet reached a consensus ~\cite{Kricheldorf2020,Lang2021,Kricheldorf2022}. Additionally, there is little literature on irreversible polymer condensation, which we also refer to as ``ligation'' henceforth in analogy with the biological process of connecting DNA segments by the enzyme ligase}.

Here we study irreversible linear polymer condensation using a combination of theory, simulations, and experiments. \rev{First, we show that irreversible polymer condensation is well captured by a modified Smoluchowski coagulation equation~\cite{smoluchowski1918versuch,ziff1980kinetics} with an additional sink term that captures ring formation}. \rev{By spanning a range of monomer concentrations $c$, we discover that above a critical $c^\dagger \simeq 0.1 c^*$ there is a ``runaway'' transition characterised by a population of chains that permanently escape cyclisation. Here $c^* = l_0/(4/3 \pi R_g^3$) denotes the overlap concentration of polymers with $l_0$ and $R_g$ the initial polymer length and radius of gyration respectively.} This transition separates a regime $(c<c^\dagger)$ in which all the chains are converted into rings at infinite time, from one ($c>c^\dagger$) in which the length of the linear chains diverges in time. \rev{The consequence of this runaway transition is that systems prepared close to $c^\dagger$ and driven out-of-equilibrium by irreversible condensation will display markedly different architectural and rheological features at large enough times}. 

\rev{Our work differs from classic and also more recent papers on polymer condensation and cyclisation~\cite{Jacobson1950,Dormidontova2004,Shimada1984, Dormidontova2004, Lang2021} because it deals with irreversible condensation while implementing subdiffusive search and cyclisation in a Smoluchowski framework and because it suggests through theory, simulations and experiments, that a runaway transition is expected beyond a critical concentration}. 
\rev{We also argue that DNA is particularly suitable to test these theories as we can readily visualise the products of ligation reactions by gel electrophoresis and distinguish linear and circular forms by treating the samples with exonuclease, as described below.} 
\rev{We conclude our paper by discussing the implications of our findings in the design of soft materials and DNA cloning.}

\section{Methods}
\subsection{\label{sec:sims}Molecular Dynamics Simulations}

\begin{figure}[t!]
    \centering
    \includegraphics[width=0.48\textwidth]{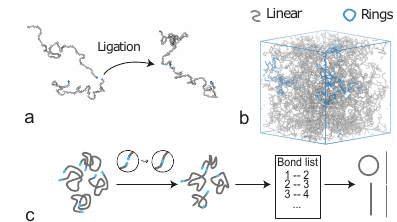}
    \caption{ \textbf{Molecular Dynamics simulations of irreversible condensation.} \textbf{a.} Sketch of a condensation (also referred to as ``ligation'') event in a Molecular Dynamics (MD) simulation. \textbf{b.} Snapshot of the simulation box with rings coloured in blue and linear chains in grey. \textbf{c.} Simulation workflow: we track the bond list to reconstruct the length and topology of the polymers and how these evolve in time. }
    \label{fig:simulation_box}
\end{figure}

We model a 6,500 bp-long linear DNA molecule as a bead-spring polymer made of $l_0=174$ beads. The total number of polymer chains is $N_c=200$. The polymers are modelled \textit{via} the Kremer-Grest model~\cite{Kremer1990}. Each bead has a diameter $\sigma=13$ nm (or $\sim 38$ bp), modelled as a truncated and shifted Lennard-Jones potential (WCA)
\begin{equation}
    U_\text{LJ}(r) = 4 \epsilon \left[ (\sigma/r)^{12} - (\sigma/r)^6 + 1/4 \right],
\end{equation}
for $r<r_c=2^{1/6}\sigma$ and 0 otherwise. Here $r$ represents the distance between beads and $\epsilon=1.0$ (in LJ units) parametrises the strength of the potential. The diameter of the bead, $\sigma$, defines the length units in our system. Consecutive beads are connected through a permanent Finite Extensible Non-linear Elastic (FENE) bond
\begin{equation}
    U_\text{FENE}(r) = - 0.5 K R_0^2 \log{\left[1 - \left( r/R_0 \right)^2\right]} 
\end{equation}
with $K=30 \epsilon/\sigma^2$ and $R_0=1.5\sigma$, which is summed to a WLC potential to yield an equilibrium bond length around $0.9 \sigma$. The bending stiffness of the polymer is controlled by a Kratky–Porod interaction 
\begin{equation}
    U_{b}(r) = \dfrac{k_BT l_p}{\sigma} (1 - \cos{\theta}),
\end{equation}
which constrains the angle ($\theta$) defined by the two tangent vectors connecting three consecutive beads along the polymer. Here, $l_{p}=4\sigma=150$ bp is the persistence length of DNA. \rev{We note that as $l_0 \gg l_p$, we are always in the flexible chain regime}. The solvent is simulated implicitly using a Langevin thermostat so that the time evolution of our system is governed by the stochastic partial differential equations 
\begin{equation}
m \ddot{\bm{r}} = -\zeta \dot{\bm{r}} - \bm{\nabla} U + \sqrt{2 k_BT \zeta} \bm{\delta}  
\end{equation}
where $\bm{r}$ is the position of a particle, $\zeta$ its friction, $m$ its mass, $U$ the sum of the interaction potentials discussed above and $\bm{\delta}$  white noise with unit variance. The diffusion timescale is $\tau_B=\zeta \sigma^2/k_BT$. The integration of the Langevin equation is done with a velocity-Verlet algorithm, using a time step $\Delta t = 0.01\tau_B$ in LAMMPS~\cite{Plimpton1995}.

Various monomer densities were considered, ranging from $10^{-2} c^*$ to $1 c^*$, where $c^* = 0.012 \sigma^{-3}$ is the monomer concentration at which the polymers start to overlap. The overlap concentration $c^*$ was measured by computing the radius of gyration $R_g$ of the polymers in equilibrium at infinite dilution. All the systems were equilibrated for a sufficient amount of time to ensure that the polymer chains have moved at least a distance equal to $R_g$.

After the equilibration step, 40 replicas of production runs were started for each number density considered. The ligation is performed stochastically and is attempted every $t_l = \tau_B$ between two end beads that are closer than $R_c = 1.1 \sigma$ using the \verb|fix bond/create| LAMMPS command. \rev{The choice of the time in between ligation attempts, $t_l$, was made so that it was much shorter than the relaxation time of the chains; in this way, the condensation process is diffusion-limited. The distance threshold $R_c$ was chosen so that the new bond created is a FENE with cutoff 1.5 $\sigma$ and to avoid unstable simulations.} The probability of successful ligation (i.e., bond formation) is set to $p_l = 0.1$. \rev{This value was chosen to avoid ``granularity'' in the stochastic condensation reaction}. If this parameter was set to 1, all the ends that can react would do so in a single time step introducing granular events in our simulations. Setting $p_l<1$ introduces some randomness that simply maps to a smaller average condensation rate. \rev{We have tested slightly different choices of these parameters and we found that the main results and qualitative behaviour of our results are not affected.} \rev{In particular, we have tested that the reactions remain diffusion-limited even with our choice of $p_l$.} A schematic representation of the simulation process is shown in Fig. \ref{fig:simulation_box}.

Once ligated, the bond formed between the polymers is irreversible and cannot be broken, therefore accounting for the formation of a covalent bond between the DNA fragments. During the ligation process snapshots of the system are taken every $10^6$ time steps on both the 3D coordinates of the beads and the bond list at those time steps. From the bond list we can, later on, reconstruct the topology of the individual polymers, i.e. if fused with others to form linear chains or if circularised. 

For the topology reconstruction, the trajectories and bond lists were analysed using our Python code (\url{https://git.ecdf.ed.ac.uk/taplab/dna-ligation.git}). The description of the algorithm can be found in Appendix \ref{sec:TR}.


\subsection{\label{sec:DSMC}The DSMC algorithm}

The modified Smoluchowski equation proposed here, (see below Eq.~\eqref{eq:smolu}), can only be solved analytically for certain forms of the condensation rate $k_1(i,j)$ and of the cyclisation rate $k_0(l)$. As our Molecular Dynamics simulations are practically limited to systems of hundreds of chains, to characterize the behaviour of larger systems we solve the Smoluchowski equation numerically employing the Direct Simulation Monte Carlo (DSMC) algorithm~\cite{garcia1987monte, liffman1992direct, kruis2000direct,tran2023fragmentation}. DSMC is a powerful stochastic method to solve differential equations such as Eq.~\eqref{eq:smolu}, and which samples the correct ligation kinetics in the limit of large system sizes. The algorithm employed here is similar to the one described in Ref.~\cite{tran2023fragmentation}, with the difference that here we do not include fragmentation, but instead, we include ring formation. The description of the algorithm also follows Ref.~\cite{tran2023fragmentation}. The starting point for the Monte Carlo algorithm is an array $\mathbf m$ of length $N_c$, each element $i$ of which contains a number $m_i$ which represents the mass/length of the chain $i$:
$$
\mathbf m = ( m_1, m_2, \dots, m_{N_c}) \, .
$$
A value of $0$ corresponds to the absence of a certain chain. Moreover, to satisfy mass conservation we ensure that $\sum_{i=1}^{N_c} m_i = N_c$ is true at any time during the simulation. Here, $N_c$ denotes the total number of polymer chains. We will also consider an analogous array $\mathbf r$ of length $N_c$ (initially empty), where we save the masses of the rings.

For an initial monodisperse condition, we set $\mathbf m_0 = ( 1,1, \dots, 1)$. After the array $\textbf m$ is initialized, we run the DSMC simulation, which consists of repeating a large number of times a Monte Carlo step (described in detail in Appendix \ref{sec:MCstep}). The execution is terminated when the system has reached a state in which there is a single linear chain and several non-reactive rings, where the only possible reaction is the cyclisation of the remaining linear chain.

\subsection{\label{sec:Exp}Experiments}

\subsubsection{Ligation Reactions with DNA}
We perform irreversible condensation on linear DNA using T4 ligase \rev{New England Biolabs (NEB).} This enzyme consumes ATP to form a covalent bond between two proximal and complementary double-stranded DNA ends. More specifically, we perform irreversible condensation on a monodisperse solution of linear, $l_0 = 6,500$ bp-long plasmid (referred to as ``1288'' plasmid here) which is converted into a linear form by using a restriction enzyme (XhoI). This linearisation step is checked on gel electrophoresis. The equilibrium radius of gyration of this linear DNA molecule is about $R_g \simeq l_p\sqrt{l_0/3l_p} \simeq 0.2$ $\mu$m (in agreement with diffusion data from Ref.~\cite{Robertson2006}). This yields an overlap concentration $c^* = 3l_0M_w/(4 N_A \pi R_g^3) \simeq 0.2$ $\mu$g/$\mu$l with $M_w=650$ g/mol the molecular weight of a DNA basepair and $N_A$ the Avogadro number. For the low DNA concentration experiments we set the sample at $0.01 c^*$, i.e. $c = 2$ $\mu$g/ml. To perform ligation we use T4 ligase (NEB, M0202L, 1U corresponds to 0.5 ng or 0.00735 pmoles of protein according to Ref.~\cite{Taylor1990}), and work at 1x T4 ligase reaction buffer concentration, which contains 1 mM ATP. To classify the topology of the DNA under ligation, we perform time-resolved gel electrophoresis. We prepare a master solution of DNA at the desired concentration, 1x ligase buffer and 2 U/$\mu$l T4 ligase. 

After adding T4 ligase, we draw aliquots at time intervals and heat-inactivate the reaction by heating the aliquot at 65$^\circ$C for 15 minutes. We then split the aliquot and treat one of the two sub-aliquots using exonuclease (RecBCD, Lucigen), an enzyme that digests linear, but not circular, DNA. Finally, we treat all aliquots with Nb.BbvCI Nickase (NEB, R0631L) to relax the supercoiled population ~\cite{Bates2005}. The resulting aliquots are run on a gel: we load 20ng of DNA from each aliquot onto a 1\% agarose gel prepared using 1x TAE buffer. A standard $\lambda$DNA - HindIII digest (NEB, N3012S) marker is also loaded. The gel is run at $\sim$ 2.5V/cm for 5 hours and post-stained with SybrGold (ThermoFisher) for 30 minutes. A Syngene G-box and Genesys software is used to image the gels. 

The combination of nickase (relaxing the DNA supercoiling) and exonuclease (fully digesting linear DNA molecules) allowed the topology of the DNA in each band to be unambiguously identified. Further, the $\lambda$DNA - HindIII digest marker confirmed the bands were of the correct size for monomer and dimer lengths. Here the terms ``monomer'' and ``dimer'' refer to a single DNA molecule and two molecules ligated, respectively. To extract the relative amount of molecules in each lane we compute, using ImageJ, the intensity of each lane and account for the fact that the band with dimers has chains that are twice as long. We then normalise against the sum of the three bands to obtain the relative fraction of chains in each population. 

\subsubsection{Microrheology}

The viscosity of the systems is measured using particle tracking microrheology. Solutions are made by mixing 8 $\mu$l of 1288 linearised plasmid at different concentrations to a final concentration in the range 2ng/$\mu$l-500ng/$\mu$l  with 1 $\mu$l of 40 U/ul T4 ligase and 1 $\mu$l of T4 ligase reaction buffer. Control solutions are prepared at the same time and in the same manner substituting additional TE for the T4 ligase. The samples are kept at room temperature on a roller for several days. The samples are then spiked with $a = 800$ nm PVP-coated polystyrene beads, pipetted and sealed onto a slide and imaged using an inverted microscope. We take a 30-minute movie and we analyse the movies using a particle tracking algorithm (trackpy~\cite{Crocker2000}) and extract the trajectories and mean squared displacements(MSD) of the tracers $\langle \Delta r^2(t) \rangle = \langle \left[ \bm{r}(t+\tau) - \bm{r}(t) \right]^2 \rangle$. Diffusion coefficients are extracted by fitting to the MSDs via MSD$=2Dt$. The viscosity is obtained using the Stokes-Einstein relation~\cite{hansen2013theory}, $\eta = k_BT/(3 \pi D a )$.

\section{Results}

\begin{figure*}[t!]
    \centering
    \includegraphics[width=0.9\textwidth]{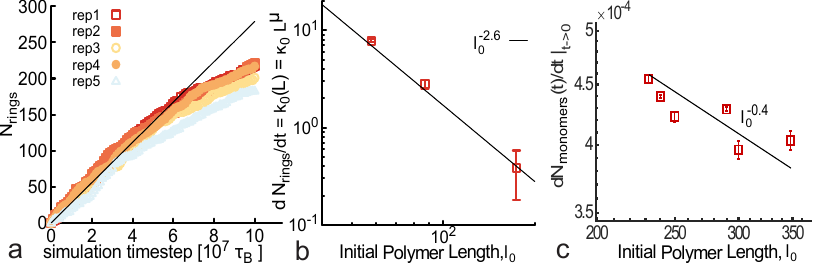}
    \caption{\textbf{Validation of cyclization and annealing rates.} \textbf{a.} Number of rings $N_\text{rings}$ as a function of time for systems initialised to $N=400$ chains with $l_0=87$ monomers each (5 independent replicas). The black line represents the numerical derivative of the average $\langle N_\text{rings} \rangle$ in the limit $t \to 0$. \textbf{b.} The numerical derivative $d N_\text{rings}/dt$ displays a power law decay with initial polymer length and with exponent $-2.6$, close to the value $-4\nu = 2.4$ for $\nu = 0.588$ (as predicted, see text Sec.~\ref{sec:SmolEq}). \textbf{c.} Value of $k_1(L)$ obtained from fitting the analytical solution of Eq.~\eqref{eq:n1} neglecting the second term, in the limit $t \to 0$. The exponent of the power law is close to the predicted value $\nu-\alpha$ (see text Sec.~\ref{sec:SmolEq}) with $\nu=0.588$ and $\alpha=1$.}
     \label{fig:rates}
\end{figure*}

\subsection{\label{sec:SmolEq}Smoluchowski equation with cyclisation}
In this result section, we first propose a modified Smoluchowski equation~\cite{smoluchowski1918versuch,ziff1980kinetics} describing polymers undergoing irreversible condensation (ligation) and cyclisation. Linear polymers undergo irreversible ligation with rate $k_1(i,j)$, with $i,j$ the polymerisation indexes of the reactants, and cyclisation with rate $k_0(q)$. The concentrations of linear polymers of polymerisation index $q$ at time $t$, $n_q(t)$, and of rings, $n_q^r(t)$, are thus governed by the following equations:
\begin{subequations}\label{eq:smolu}
\begin{empheq}[left=\empheqlbrace]{align}
\dot n_q(t) & = \frac 1 2 \sum_{ij; i+j=q} k_1(i,j) n_i(t) n_j(t)  + \notag \\ & -  n_q(t)\sum_{i=1}^\infty k_1(q,i) n_i(t) - k_0(q) n_q(t)\label{eq:smolu1}
\\
\dot n_q^r(t) & = k_0(q) n_q(t) \label{eq:smolu2}\,  .
\end{empheq}
\end{subequations}
Once a linear chain undergoes cyclisation, it becomes a ring and cannot undergo ligation anymore, as the reactions are assumed to be irreversible. The kinetics is also constrained by the requirement that the total mass is conserved:
\begin{equation}
 \sum_{q=1}^\infty q [n_q(t)+n_q^r(t)] = M/V = n \quad \forall t \, ,
 \end{equation}
where $M$ is the total number of monomers and $V$ is the system's volume. Assuming that the reaction takes place on a time scale larger than the Rouse relaxation time, the length-dependence of the annealing rate is~\cite{DeGennes1982a,grosberg1982polymeric}
\begin{align}
    k_1(i,j) & = \tilde \kappa_1 (D_i + D_j) (R_i + R_j) \\
& = \kappa_1 \left(i^{-\alpha} + j^{-\alpha} \right) \left( i^{\nu} + j^{\nu} \right) \, ,
    \label{eq:smolu_rates}
\end{align}
\noindent
where $l = i l_0$ is the length of a polymer with a degree of polymerisation $i$ and $l_0$ is the initial polymer length, so that the chain's radius of gyration is $R_i = l_0 i^\nu$. In Eq.~\eqref{eq:smolu_rates}, $\tilde \kappa_1$ is a dimensionless constant and $\kappa_1$ is a constant that depends on temperature and the viscous friction of the solvent $\zeta$. For example, in the Rouse model~\cite{Doi1988} $D_i=k_B T/(\zeta l_0 i)$ ($\alpha=1$) and thus $\kappa_1 = \tilde \kappa_1 k_B T/\zeta$.

\rev{This condensation rate captures the diffusion-controlled search process~\cite{DeGennes1982a,grosberg1982polymeric}.} The cyclisation rate is taken to be $k_0(q) = \kappa_0 q^{\mu}$, where $\mu=-4\nu$. Note that this is different from the classic Shimada-Yamakawa theory~\cite{Shimada1984,Rosa2010} which would predict $\mu=-3\nu$ at lengths larger than $l_p$ because we (i) are out-of-equilibrium and (ii) account for the subdiffusion of the polymer end within the volume of the coil. 

In equilibrium, the looping probability of a chain is given by the Shimada-Yamakawa formula~\cite{Shimada1984,Rosa2008}. For $l \gg l_p$ the looping probability of a polymer decays as $P(l) \sim l^{\mu}$ with $\mu = -3 \nu$. This looping probability also holds for an irreversible, non-equilibrium scenario if the process is reaction-limited. This is because the chain ends would have the time to explore many conformations and to diffuse the whole volume of the chain, $V \sim l^{3 \nu}$, before reacting (as it would happen in equilibrium). 
In a diffusion-limited, irreversible ligation process, one should instead compute the time it takes for an end to diffuse over a certain distance $\xi$. The dynamics of the end is described by the Rouse model~\cite{Doi1988} so that $\xi = b [k_B T t/(\zeta b^2)]^{1/4}$, where $b$ is the size of a Kuhn monomer. Then, setting $\xi = R$ (the size of the polymer coil) one obtains $(R/b)^4 = k_B T t/(\zeta b^2)$, which implies $k_0 \sim t^{-1} \sim R^{-4} \sim l^{-4\nu}$. So considering $\mu = -4 \nu$ effectively takes into account the fact that the chain ends are performing a sub-diffusive search process within the polymer coil, as expected for Rouse dynamics. 

\rev{We have verified that the rate of cyclisation scales as the length of the chain to the power $-4 \nu$ by measuring the rate at which rings are produced for different lengths of the linear chains (Fig.~\ref{fig:rates}a-b). We have done this by changing the initial length $l_0$ and by running short simulations, in turn assuming that the system has had no time to create dimers, trimers, etc. and by measuring the number of rings formed. We have observed that the rate of ring formation at early times $\dot N_\text{rings} \sim l_0^{-2.6}$ which is close to the expected $\mu = - 4\nu = 2.4$ with $\nu=0.588$. Thus, both theory and simulations suggest that the \emph{diffusion limited, irreversible} looping probability of a polymer scales with its length as $l^{-4\nu}$.}

\rev{To validate the functional form used for the condensation rate $k_1(i,j)$ (Eq.~\eqref{eq:smolu_rates}), we solve the Smoluchowski equation in the limit of small concentration and short times, where only monomer, dimer and monomer ring populations are assumed to be present (see next Section). In Fig.~\ref{fig:rates}c we plot the condensation rate $k_1$ as a function of different initial polymer lengths obtained by fitting the analytical solution of Eq.~\eqref{eq:n1} (see below) to the monomer chains population omitting the second term since no rings were present in these conditions and at early times. From this quantity, we fit a power law $l^{\nu-\alpha}$ with $\nu=0.588$ and find $\alpha \simeq 1$ yields a good fit to the simulated data. This validates de Gennes' hypothesis for the functional form of the condensation rate (Eq.~\eqref{eq:smolu_rates}) and our choices for $\nu$ and $\alpha$.}

In our experiments and simulations, we typically track the mean length of the polymers as a function of time
\begin{equation}
\langle l (t) \rangle  = l_0 \dfrac{\sum_{i=1}^{\infty} i (n_i(t) + n_i^r(t))}{\sum_{i=1}^{\infty} (n_i(t) + n_i^r(t))} \, ,
\end{equation} 
and we fit this observable with the numerical solution of the full Smoluchowski equation, Eqs.~\eqref{eq:smolu1}-\eqref{eq:smolu2}). This is practically implemented in a MATLAB code. The numerical evaluation of the system is iterated to find the best free parameters $\kappa_0$ and $\kappa_1$ that fit the mean length $\langle l (t) \rangle$ obtained from simulations or experiments. 
\rev{The fit is done using the nonlinear least squares MATLAB function {\tt lsqcurvefit}. The rate of ring formation $\kappa_0$ and the rate of linear chains formation $\kappa_1$ are extracted from this fit by considering 40 independent replicas, allowing us to obtain the error on the fit parameters. For experiments, we typically average over 3 independent replicas. The numerical and fitting algorithms are described in detail in Appendix \ref{sec:numIntSmol}.}

\subsubsection{Time-dependence of the mean length: dilute regime}

\rev{At short times and in the dilute regime, we can assume that the formation of rings and short $n$-mers is more favourable. This assumption is valid in the experiments whenever only linear monomers, dimers and monomer rings are visible in the gel electrophoresis after ligation. In more dense solutions the presence of rings consisting of more than two monomer chains will be present and is observed in our simulations.} 
Under very dilute conditions, we can thus assume that only monomers, dimers and monomer rings are present. Denoting the number density of monomer rings, linear monomers and dimers as $n_1^r, n_1$ and $n_2$, respectively, the Smoluchowski equations describing the system take the form
\begin{subequations}\label{eq:smolu_3by3}
\begin{empheq}[left=\empheqlbrace]{align}
&\dfrac{d n_1^r(t)}{dt} = k_0(1) n_1(t) \label{eq:n1r} \\ 
&\dfrac{d n_1(t)}{dt} = - k_1(1,1) n_1^2(t) - k_0(1) n_1(t) \label{eq:n1} \\ 
&\dfrac{d n_2(t)}{dt} =  \dfrac{1}{2} k_1(1,1) n_1^2(t) \label{eq:n2}\, .
\end{empheq}
\end{subequations}
We solve Eq.~\eqref{eq:n1} neglecting the second term as $n_1^2 \ll1$ in the infinite dilution limit:
\begin{equation}
n_{1}(t) = n_1(0)e^{-k_0(1)t}    
\label{eq:smol_closed_mon}
\end{equation}
The concentration of monomer rings is thus
\begin{equation}
\dfrac{d n_1^r(t)}{dt} = k_0  n_1(0) e^{-k_0(1)t} \, 
,\end{equation}
which yields
\begin{equation}
n_1^r(t) = n_1(0) [1-e^{-k_0(1)t}] \, .
\label{eq:smol_closed_rings_mon}
\end{equation}
Substituting in Eq.~\eqref{eq:n2}, we get
\begin{equation}
\dfrac{d n_2(t)}{dt} =\dfrac{1}{2} k_1(1,1) {n_1}^2(t) =  \dfrac{1}{2} k_1(1,1) \left[n_1(0) e^{-k_0(1)t}\right]^2 \, ,
\end{equation}
from which one obtains 
\begin{equation}
n_2(t) = \dfrac{k_1(1,1)}{4 k_0(1)} {n_1}^2(0)  \left[1-e^{-2k_0(1)t}\right] \, .  
\label{eq:smol_closed}
\end{equation}
Assuming these three are the only contributions to the system, the mean length is then given by the following relation
\begin{align}
\langle l(t) \rangle &= \dfrac{l_0 n_1(t) + l_0 n_1^r(t) + 2 l_0 n_2(t)}{n_1(t) + n_1^r(t) + n_2(t)} =  \notag \\ 
 & = l_0 \dfrac{n_1(t) + n_1^r(t) + 2 n_2(t)}{n_1(t) + n_1^r(t) + n_2(t)} \, .
\end{align}
In denser solutions, where the population is more polydisperse, the Smoluchowski equation cannot be solved analytically and we refer to the next Section for a scaling prediction and to Sec. \ref{sec:perturbation} for a perturbative approach in the limit of small cyclisation rate.

As mentioned above, we validate de Gennes' equation for the condensation rate (Eq.~\eqref{eq:smolu_rates}) by running short simulations at very high dilution. We then fitted the change in number of ring monomers with the closed solutions Eqs.(\eqref{eq:smol_closed_rings_mon}) for different values of initial polymer length $l_0$. Similarly, we fit the solution of Eq.~\eqref{eq:n1} without the ring term to the population of monomers. From these data, we validate the scaling of the rates $k_0(l_0) = \kappa_0 l_0^{-\mu}$ and $k_1(l_0,l_0) = \kappa_1 l_0^{\alpha-\nu}$ as a function of length $l_0$ (Fig.~\ref{fig:rates}b-c). 
 
\subsubsection{\label{sec:tdep_conc}Time-dependence of the mean length: concentrated regime}

Here we give scaling arguments for the solution of the Smoluchowski equation in the concentrated limit, with the assumption that ring formation is negligible. At the mean-field level, we can make the simplifying assumption that the system can be described by a \textit{single characteristic length scale} $l$~\cite{sorichetti2023transverse}. Under this assumption, the annealing rate scales as
\begin{equation}
    k_1(l) \sim D R \sim l^{\nu - \alpha} \, . 
    \label{eq:k1_scaling}
\end{equation}
The total polymer density $n$ thus follows $\dot n = - k_1 (l) n^2$, so that from the dimensional analysis the time evolution of the characteristic length is~\cite{vandongen1984kinetics,vandongen1985dynamic}
\begin{equation}
    l(t) \sim t^{1/(1+\alpha-\nu)} \sim t^{ 1 /(1-\lambda)} \equiv  t^\gamma \, ,
    \label{eq:l_vs_t}
\end{equation}
with $\lambda = \nu - \alpha$.  For Rouse dynamics, one has $\alpha=1$, whereas $\alpha=2$ for reptation~\cite{Doi1988b}. The Flory exponent has value $\nu=1/2$ for ideal chains and $\nu=0.588$ for self-avoiding chains~\cite{Doi1988b}. Assuming concentrations above overlap but still far from the melt concentration (for which one would have ideal chain statistics and $\alpha=1/2$), we can assume $\nu=0.588$, so that $\gamma\simeq0.7$ if the system is unentangled and $\gamma\simeq 0.4$ in the presence of entanglement. We note, however, that using Eq.~\eqref{eq:k1_scaling} in the presence of entanglements is only valid for times longer than the reptation time $\tau_R \sim l^3$~\cite{DeGennes1982}.

\begin{figure}[t!]
    \centering
    \includegraphics[width=0.48\textwidth]{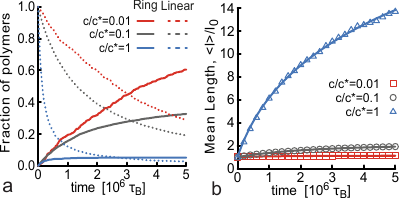}
    \caption{\textbf{Results from MD simulations.} \textbf{a.} Number fraction of linear (dotted line) and ring (solid line) polymers during the condensation process for different concentrations $c$ (averaged across 40 independent replicas). \textbf{b.} Number averaged polymer length for the simulations in \textbf{a} as a function of time (symbols), and fitted (solid line) with a numerical solution of Eq.~\eqref{eq:smolu} (see Sec.~\ref{sec:SmolEq} for details).}
     \label{fig:looping_sims}
\end{figure}
 
\subsection{Linear DNA condensation}

\subsubsection{Simulations}

We first simulate linear condensation using Molecular Dynamics. As detailed in the Methods section, we simulate polymers with $N=174$ beads of size $\sigma \sim 38$ bp and persistence length $l_p=4 \sigma = 150$ bp. These polymers are thus designed to coarse-grain 6.5 kb-long DNA plasmids which will be employed in experiments (see next section).
During the simulation, we take snapshots of the system and record the list of bonds to reconstruct the topology of the polymers (see Fig.~\ref{fig:simulation_box}). Over the simulation time, the number of initial linear chains decreases due to the formation of (i) longer linear polymers or (ii) circular chains (Fig.~\ref{fig:looping_sims}a). Additionally, lower monomer concentrations $c$ promote the formation of more rings at large times and a slower decrease of the linear species. We also note that (i) the number fraction of rings converges to a finite value at large time, and that (ii) while the number of linear chains appears to go to zero, their mean length increases (Fig. \ref{fig:looping_sims}b). Accordingly, the (number) average length of polymers grows more quickly for larger $c$ (Fig.~\ref{fig:looping_sims}b). Thus, we conclude that loop formation competes with the growth of the chains, and that cyclisation is dominant in dilute systems.  \rev{Interestingly, the curves of the mean length $\langle l (t)\rangle$ can be fitted extremely well by the numerical solution of the Smoluchowski equation Eq.~\eqref{eq:smolu} (Fig.~\ref{fig:looping_sims}b).}

\subsubsection{Experiments}

\begin{figure}[t!]
\centering
     \includegraphics[width=0.48\textwidth]{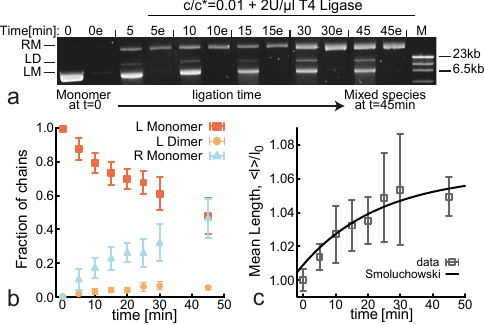}
     \caption{\textbf{Experiments of DNA condensation at low concentration.} \textbf{a.} Time-resolved gel electrophoresis during ligation of a 2 ng/$\mu$l ($c/c^* = 0.01$) solution of 6.5 kb linearised plasmid. The lanes marked with `e' are treated with exonuclease to remove linear DNA; `RM' indicates ring monomers, `LD' linear dimers and `LM' linear monomers. The term monomer here refers to a single DNA molecule. The last column is $\lambda$-HindIII marker as a reference of DNA fragment lengths. The numbers at the top represent the minutes for which the solution was incubated with ligase enzyme. \textbf{b.} Number fraction of polymers in ring and linear (monomer and dimer) topologies obtained from image analysis of the gel in \textbf{a}. \textbf{c.} Number averaged length calculated from 3 independent gels at 2 ng/$\mu$l (symbols) and associated fit (solid line) using Eq.~\eqref{eq:smolu} numerically solved as described in Sec.~\ref{sec:SmolEq}.}
     \label{fig:looping_exp}
 \end{figure}

\rev{As described in the Methods Section, we can perform DNA condensation using solutions of linearised DNA plasmids, mixed with ATP and DNA T4 ligase. We then perform a time-resolved experiment, where we draw aliquots from a master reaction at given time points from the addition of the T4 ligase.} By running the aliquots on agarose gels we can visualise and compute the fraction of molecules in the linear and ring, monomeric, dimeric, etc. states. Fig.~\ref{fig:looping_exp}a reports a picture of one such gel, displaying a single band of monomeric linear DNA (as it disappears after exonuclease treatment) at $t=0$, evolving into three bands, one of which is exonuclease resistant (a monomer ring) at larger times. In Fig.~\ref{fig:looping_exp}b we plot the relative abundance of these populations, from which we obtain the number average molecular length $\langle l(t) \rangle$ (Fig.~\ref{fig:looping_exp}c).

 \subsubsection{Dimensionless topological parameter}
Since we initialise our simulations and experiments below entanglement conditions we \rev{fix $\alpha=1$ as expected for Rouse dynamics and $\nu=0.588$ as expected for self-avoiding polymers~\cite{Doi1988} (we verified these exponents through direct MD simulations in Fig.\ref{fig:rates}a-c).} In general, the Smoluchowski coagulation equation (Eq.~\eqref{eq:smolu}) is then solved numerically to fit the data of mean length versus time, $\langle l (t) \rangle$, obtained in simulations and experiments via the free parameters $\kappa_1$ and $\kappa_0$. A key number in our system is the ratio of the rates at which polymers are condensed $\kappa_1$, and the one at which rings are formed $\kappa_0$. We thus define a dimensionless ``topological parameter'' $\kappa \equiv 2 \kappa_0/(n_0 \kappa_1)$, where $n_0$ is the number density of monomeric chains of length $l_0$ at the start of the simulation or experiment.

Albeit related to the classic j-factor employed in DNA looping~\cite{Shimada1984,VafabakhshRezaHa2012}, our topological parameter is more naturally interpreted as the number of rings formed for every two linear chains that are fused together. Intuitively, this number determines the final topological composition of the system. At $\kappa \gg 1$, we expect the final state of the system to be dominated by rings, while for $\kappa \ll 1$ to be dominated by linear chains. Importantly, since $k_0 \sim \langle l(t) \rangle^{-4\nu}$ the probability of ring formation decreases in time as the average length of the linear chains increases. Accordingly, and even though our system has a ring-only irreversible absorbing state, we conjecture that the strongly decreasing looping probability may effectively yield a very long time-transient in which the system is dominated by entangled linear chains with circular contaminants (see below for more simulations on this). 

\begin{figure}[t!]
    \centering
    \includegraphics[width=0.30\textwidth]{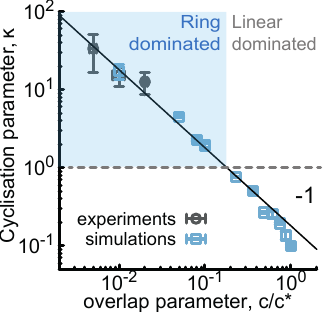}
    \caption{\textbf{The dimensionless topological parameter.} Dimensionless topological (also ``cyclisation'') parameter $\kappa = 2 \kappa_0/(n_0 \kappa_1)$ as a function of $c/c^*$. This is obtained by fitting simulations and experimental curves $\langle l(t) \rangle$ with the numerical solution of the Smoluchowski equation with $\kappa_0$ and $\kappa_1$ as free parameters. There are no other free parameters in our model. The scaling $\kappa \sim (c/c^*)^{-1}$ is consistent with $\kappa_0$ and $\kappa_1$ being independent of concentration, and with this assumption breaking down near $c/c^* \simeq 1$ where three-body interactions become important.}
    \label{fig:runaway_1}
\end{figure}

Importantly, we expect the Smoluchowski equation to be valid only in the limit in which three-body interactions are negligible, the values of $\kappa_0$ and $\kappa_1$ should be independent on concentration only when $c$ is small enough. By plotting $\kappa \equiv 2 \kappa_0/(n_0 \kappa_1)$ as a function of $c/c^*$ (where $c^*$  is computed at the beginning of the simulation or experiment) we show that $\kappa$ scales as $n_0^{-1} \sim (c/c^*)^{-1}$ in both simulations and experiments until $c \simeq c^*$ where it starts to deviate (Fig.~\ref{fig:runaway_1}); this confirms that the Smoluchowski approximation is valid in this range of concentrations. Importantly, in Fig.~\ref{fig:runaway_1} we also identify the crossover value $\kappa=1$ (at which the initial cyclisation rate is larger than the dimerisation rate) around $c/c^* \simeq 0.1-0.2$. We note that the agreement between simulations and experiments is excellent for small $c/c^*$. However, quantitative analysis of gel electrophoresis images at larger $c/c^*$ is challenging due to the poor separation of multimeric bands. 

\begin{figure*}[t!]
    \centering
\includegraphics[width=1.0\textwidth]{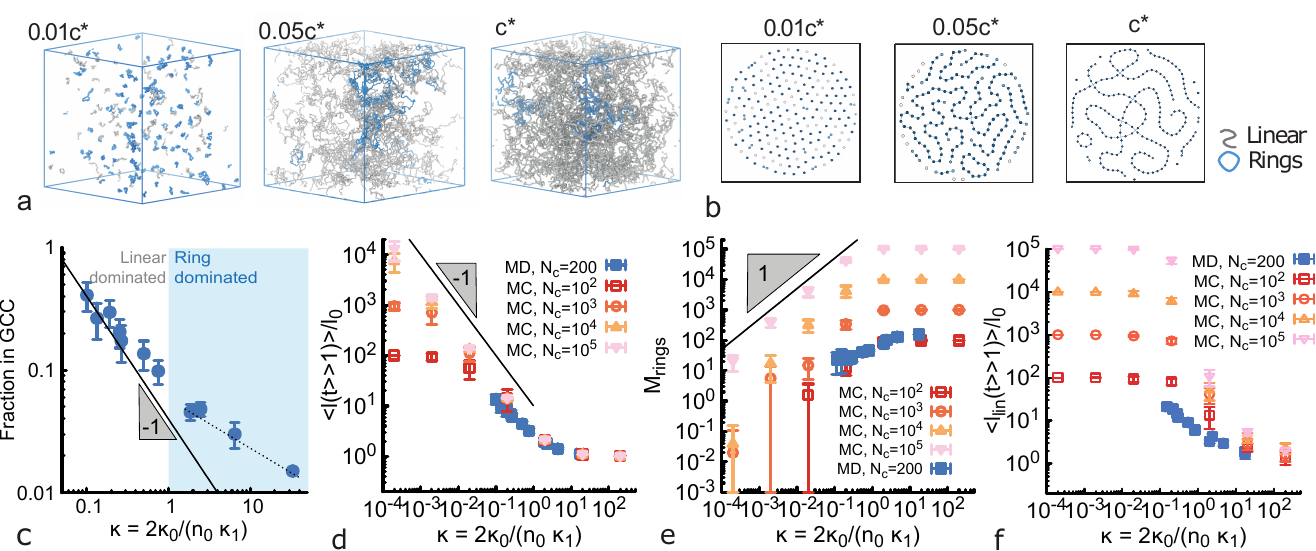}
    \caption{\textbf{Runaway Transition.} \textbf{a.} Snapshots of MD simulations with initially $N_c=200$ chains: the blue polymers represent rings of any length, grey polymers linear chains. \textbf{b.} In the corresponding graph representations, each circle represents a single chain and its colour represents the number of chains connected to it: 0 (grey), 1 (cyan), or 2 (blue). For example: a linear monomer chain is grey, a ring monomer is blue, a linear dimer has two cyan nodes, and a dimer ring has two blue nodes.
    \textbf{c.} Fraction of monomers in the giant connected component of the system as a function of $\kappa$. \textbf{d.} Average polymer length (including both linear and ring) at a large time as a function of $\kappa$. \textbf{e.} Average mass of rings $M_\text{rings} = N_\text{rings} \times \langle l_\text{rings}(t \gg 1) \rangle$ at large simulation time. Note that some simulations may not display any rings, thus bringing the average $M_\text{rings}$ below 1. \textbf{f.} Average mass of linear chains at large simulation time as a function of $\kappa$. MC = “Monte Carlo”; MD=“Molecular Dynamics”. }
     \label{fig:MC_MD}
\end{figure*}

\subsection{\label{sec:runaway}Runaway Transition} The results in Fig.~\ref{fig:looping_sims} suggest that at large $c/c^*$ the chains tend to grow longer, and cyclisation is suppressed; at the same time, the density of reactive ends and the speed of spatial exploration of the chains become smaller, thus suppressing dimerisation. Due to this kinetic competition, we ask whether the system can truly display a ``runaway'' phase, defined as a regime where at least one chain permanently escapes cyclisation and its length diverges in time. One way to address this question is to look at the number of chains that belong to the longest chain in the system, and how this quantity changes in time. 
By using a graph representation of our simulations (see Fig.~\ref{fig:MC_MD}a-b) we can compute the fraction of chains (nodes) that belong to the giant connected component (GCC), i.e. the largest cluster of connected monomer chains (Fig.~\ref{fig:MC_MD}b). In Fig.~\ref{fig:MC_MD}a one can visually appreciate that at large reaction time, rings (blue) are abundant at low $c/c^*$ while linear chains (grey) are more abundant at large $c/c^*$. These systems display a qualitatively different graph topology (Fig.~\ref{fig:MC_MD}b). At small $c/c^*$ (large $\kappa$) the network of monomers is mostly disconnected; accordingly, even when the fraction of unreacted bonds goes to 0 at $t \to \infty$, the average length of the polymers does not diverge. On the contrary, at larger $c/c^*$ we observe only a few rings and some very long chains that are connecting most of the nodes in the system. Overall, the graph appears much more connected and approaching percolation, i.e. where most of the nodes belong to the GCC, whose size grows with the size of the system (see Fig.~\ref{fig:MC_MD}c).

\subsubsection{\label{sec:perturbation}Calculation of mass converted into rings at infinite time}

Although our simulations support the notion that small values of $\kappa$ will result in linear chains of increasing lengths and vanishing cyclisation rate, they are fundamentally limited to finite-size systems where the cyclisation rate of the largest chain never rigorously goes to 0. To estimate the amount of mass that is converted into rings at long times, we do a perturbative calculation valid in the limit of small $\kappa$.  We start from the continuum Smoluchowski equation: 
\begin{align}
 & \dfrac{d n_l (t)}{dt} = \dfrac{1}{2} \int_0^l K(y,l-y)n_y(t) n_{l-y}(t) dy  + \notag \\ 
 & -  \int_0^\infty K(y,l) n_l(t) n_y(t) dy - \dfrac{1}{2} \kappa l^{\mu} n_l(t) \, ,
 \label{eq:smol_cont}
\end{align}

We define $K \equiv  k_1/\kappa_1$ which is thus a scaling function such that $K(a i,a l) \sim a^{\lambda} k_1(i,l)$ where $\lambda = \nu - \alpha$~\cite{meakin1988scaling}. We now treat $\kappa_0$ perturbatively, starting with $\kappa_0=0$. In this case, there is no mass lost into rings and we can thus write a conservation law 
\begin{equation}
    \int_0^\infty l n_l(t) dl = 1 \quad \forall t \, . 
\end{equation}
Even for $\kappa$ non-zero, we assume the loss of mass to cyclisation remains finite and of order $\kappa$. We will check the self-consistency of this assumption below. Using the mass conservation and Eq.~\eqref{eq:smol_cont} we can write the following scaling relations: $l^2 n = 1$, $n t^{-1} = l^{1+\lambda} n^2$. We therefore obtain: 
\begin{equation}
    l \sim t^{1/(1 - \lambda)} \, .
\end{equation}
which is the same as Eq.~\eqref{eq:l_vs_t}. Note that we must have $\lambda < 1$ for the average length of polymers to increase over time. We can also write the density distribution as 
\begin{equation}
n \sim t^{-2/(1 - \lambda)}   
\end{equation}
which in the limit of long times or large lengths may be written as 
\begin{equation}
n_l(t) \sim t^{-2/(1 - \lambda)} \mathcal{G}\left(\dfrac{l}{t^{1/(1-\lambda)}}\right)    \label{eq:rho_l}
\end{equation}
where $\mathcal{G}$ is a scaling function that only depends on the ratio $l/t^{1/(1-\lambda)}$.

We now introduce the ring length distribution $n_l^r(t)$ and its evolution equation as 
\begin{equation}
    \dfrac{d n_l^r(t)}{dt} = 2 \kappa_0 l^{\mu} n_l(t)  \, .
\end{equation}
Since at time $t=0$ there are no rings, we can then write 
\begin{equation}
    n_l^r (t\to \infty) = 2 \kappa_0 l^{\mu} \int_0^{\infty} n_l(t) dt \, .
\end{equation}
We can plug in the result we obtained for the distribution of length of linear chains Eq.~\eqref{eq:rho_l} to yield
\begin{align}
    n_l^r( t\to \infty) & = 2 \kappa_0 l^{\mu} \int_0^{\infty} t^{-2/(1 - \lambda)} \mathcal{G}\left(\dfrac{l}{t^{1/(1-\lambda)}}\right) dt \notag \\
    & =  2 \kappa_0 l^{\mu} (1-\lambda) l^{-(1+\lambda)}\int_{0}^\infty x^{-\lambda} \mathcal{G}(x) dx  
\end{align}
where we defined $x = l/t^{1/(1-\lambda)}$. Thus, the number density of polymers that are converted into rings over infinite time is 
\begin{equation}
    n_l^{r\infty} = 2 \kappa_0(1-\lambda) l^{\mu - 1 - \lambda} \int_0^\infty x^{\lambda} \mathcal{G}(x) dx \, .
    \label{eq:mass_ring_infinity}
\end{equation}
Since $\lambda < 1$ and assuming the $\mathcal{G}(x) = \mathcal{O}(1)$ when $x \to 0$, the integral converges at 0. For convergence of this integral at $\infty$ we also require that the scaling function decays faster than $x^{\lambda - 1}$.

Assuming this functional form for the distribution of ring lengths at infinite time, we now compute the total average mass transformed into rings at infinite time as 
\begin{align}
    M_\text{rings}^{\infty} &= \int_1^{\infty} l n_l^{r\infty}(t) dl \notag \\
 &  = 2 \kappa_0 (1 -\lambda) \int_0^{\infty} x^{-\lambda} \mathcal{G}(x) dx \int_1^{\infty} l^{\mu - \lambda} dl \, .
\end{align}
The convergence of this integral requires that $\lambda - \mu > 1$ and in this case we get 
\begin{equation}
     M_\text{rings}^{\infty} = 2 \kappa_0 \dfrac{1- \lambda}{\lambda -\mu - 1} \int_0^\infty  x^{-\lambda} \mathcal{G}(x) dx \, .
\end{equation}
From this equation, we see that the fraction of mass in rings at infinite time $M_\text{rings}^{\infty}/M_0$ converges to a finite value proportional to $\kappa_0$ (and hence $<1$ at small $\kappa_0$). 

With this calculation, we have thus shown that at small enough but non-zero $\kappa_0$, the fraction of mass turning into rings is finite if $\lambda - \mu = \nu - \alpha + 4\nu > 1$ or $\nu > \alpha/5$ which is valid for any type of polymer in the non-entangled ($\alpha=1$) regime. 
This implies that in this regime we expect the cyclisation probability to decay fast enough and cannot prevent the runaway of the $M_0 -  M_\text{rings}^{\infty}$ mass into linear chains that keep growing in time.

Consistently with this, in both MD and MC simulations, we never observe the formation of rings larger than 10 initial monomers. As shown here using asymptotic theory the mass fraction of linear polymers goes to a finite limit at $t\rightarrow\infty$ in a thermodynamic system. We find that the key condition to ensure the existence of runaway transition is that the cyclisation rate $k_0 = \kappa_0 l^{\mu}$ decays strongly enough. More specifically, we require the exponent $\mu$ to be $\mu = - 4\nu < -4\alpha/5$ or $\nu > \alpha/5$. This condition is always met in the Rouse unentangled ($\alpha=1$) regime, 
provided that the polymers are not fully collapsed ($\nu=1/3$). This argument establishes the existence of a runaway transition in the limit of large time and at large enough concentrations $c/c^*$. 

\subsubsection{\label{sec:DSMCresults} Direct Simulation Monte Carlo  simulations of irreversible condensation}

\rev{To formally address the existence of a true runaway transition in the thermodynamic limit, we compute the fraction of monomers belonging to linear species in systems of increasing size. To perform this calculation, we employ Direct Simulation Monte Carlo~\cite{garcia1987monte, liffman1992direct, kruis2000direct, tran2023fragmentation} to solve the Smoluchowski equation in systems with up to $10^5$ chains. We run the DSMC code until it has reacted all ends apart from 2 and compute the average length of the \emph{linear} population of chains, $\langle l_{lin}(t \gg 1) \rangle$.} As shown in Fig.~\ref{fig:MC_MD}c, our MD simulations show that at $\kappa=1$ the GCC displays a change in scaling, growing as $GCC \sim \kappa^{-1} \sim c/c^*$ as $\kappa \to 0$ suggesting that a qualitative change in behaviour takes place around $\kappa \simeq 1$. In Fig.~\ref{fig:MC_MD}d, we also plot the number averaged chain length at an arbitrarily large time when the DSMC code has evolved the system as long as possible and has generated only a single linear chain. Fig.\ref{fig:MC_MD}d suggests that the linear-dominated regime ($\kappa < 1 $) displays an average polymer length at a large reaction time that scales as $\langle l(t \gg 1) \rangle/l_0 \sim \kappa^{-1} \sim c/c^*$. Additionally, the fraction of mass ``lost'' in forming rings grows as $M_{rings} \sim \kappa$ and is thus negligible for small enough $\kappa$ (Fig.~\ref{fig:MC_MD}e). Finally, as shown in Fig.~\ref{fig:MC_MD}f, the mean length of the linear chains $\langle l_{lin}(t \gg 1) \rangle$ displays a plateau for $\kappa \lesssim 1$, which grows with the system size, strongly indicating a true runaway transition at the critical value $\kappa \simeq 1$ or $c/c^* \simeq 0.1-0.2$.


\subsection{\label{sec:Dyn}Dynamics and Rheology} 
\begin{figure}[h]
    \centering
\includegraphics[width=0.48\textwidth]{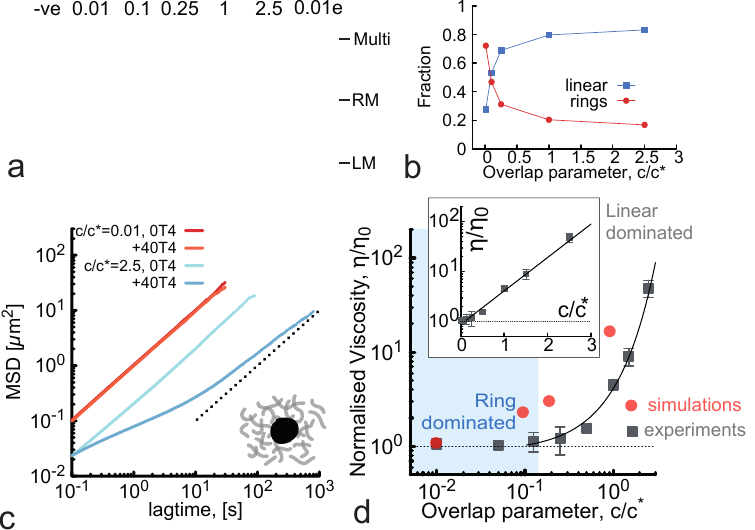}
    \caption{ \textbf{Rheological consequences of the runaway transition.} \textbf{a.} Time-resolved gel electrophoresis during ligation of $c/c^* = 0.01, 0.1, 0.25, 1$ and $2.5$ solution of 6.5 kb linearised plasmid. The lanes marked with `e' are treated with exonuclease to remove linear DNA; `RM' indicates ring monomers, `LM' linear monomers and `Multi' various lengths of linear structures. The numbers at the top represent the concentrations of DNA. \textbf{b.} Fraction of linear and ring molecules of any size as a function of overlap parameter $c/c^*$.
    \textbf{c.} Mean squared displacements (MSDs) of 800 nm PVP-coated polystyrene beads diffusing in solutions of DNA. We compare the MSDs in solutions treated for 1 week either with 40U T4 ligase or buffer (control). \textbf{d.} Viscosity of the ligated solutions $\eta$, normalised by the viscosity of the control $\eta_0$ \emph{vs} $\eta/\eta_0$ of the ligated solutions in simulations. The inset shows the same plot in a log-linear scale to highlight the exponential increase.}
    \label{fig:dynamics}
\end{figure}

To test the consequences of the runaway transition on the dynamics and rheology of the system, we perform microrheology experiments and compute dynamics in MD simulations. \rev{DNA microrheology is well established and the effects of DNA concentration, length and topology on microrheology have been studied in the past~\cite{Mason1997,Zhu2008,Krajina2017,Tanoguchi2018,Smrek2021,Michieletto2022natcomm,Fosado2023}}. Here, we perform microrheology by tracking $800$ nm PVP-coated polystyrene beads added in a solution of DNA that has been treated with either 40U T4 ligase for a week (and thus to full extent of reaction) or with buffer for a week (control) at different initial concentrations. \rev{We ran a small aliquot of the samples in a gel and observed that indeed at $c/c^* \simeq 0.1$ the fraction of linear chains overcome the rings at large times (Fig.~\ref{fig:dynamics}a-b).} 

At low concentrations, our microrheology shows that the MSD of the tracer particles is unaffected by DNA ligation (Fig.~\ref{fig:dynamics}c). On the contrary, for $c/c^* \geq 0.1$, we find that the MSDs of the tracers in the ligated systems are much slower and display a stronger subdiffusive behaviour than the control (Fig.~\ref{fig:dynamics}c). From the MSD, we extract the large-time diffusion coefficient $D$ of the tracers and the effective viscosity of the sample via the Stokes-Einstein equation~\cite{hansen2013theory}. The plot of the normalised viscosity (Fig.~\ref{fig:dynamics}d) suggests that a dynamical transition takes place around $c/c^* = 0.1 - 0.2$ (or $\kappa \simeq 1 - 2$) which matches the structural runaway transition seen before (Fig.~\ref{fig:MC_MD}). After the transition, the viscosity increases exponentially with the concentration (see inset of Fig.~\ref{fig:dynamics}d). This suggests a relaxation process dominated by end-retraction~\cite{Doi1988}, possibly due to the threading of very long linear chains through small rings~\cite{Roovers1988,Kapnistos2008,Halverson2012,Zhou2021,Parisi2020} or pseudo-knotted parts of their own extremely long  contour~\cite{Michieletto2014self,Soh2019a}. We note that, especially at large $c/c^*$, the ligated solution is extremely elastic and the passive tracers do not display a freely diffusive behaviour even after a lag time of ten minutes. We thus argue that the reported $\eta/\eta_0$ may be lower bounds at large $c/c^*$, which would render the transition even more dramatic. All this implies that, intriguingly, near the transition $c/c^* \simeq 0.1$, systems prepared at similar concentrations may display extremely different rheology at large condensation times. To further support the existence of a qualitative change in the dynamics, we compute the values of viscosity obtained in MD simulations through the diffusion coefficient of the centre of mass of chains that have been ligated for long time at different initial concentrations (see red circles in Fig.~\ref{fig:dynamics}d). \rev{One can appreciate that our simulations also suggest a qualitative difference in dynamics for $c/c^* \geq 0.1$, albeit the transition appears less dramatic than in experiments; we argue that this may be due to finite size effects present in MD simulations.}

\section{Conclusion}
We have studied a system of linear polymers undergoing irreversible condensation in competition with cyclisation. We have shown that the key adimensional parameter controlling growth kinetics is $\kappa = 2 \kappa_0/(n_0 \kappa_1)$; naturally interpreted as the number of rings formed for any one dimerisation. At large concentrations (or $\kappa<1$) dimerisation is kinetically favoured and drives the growth of linear chains. While growth disfavours cyclisation, it also reduces the number of available reactive ends and the annealing rate of the chains (see Eq.~\eqref{eq:smolu_rates}), disfavouring further growth. Despite this, we discover that the net result of this kinetic competition is a runaway transition for $\kappa<1$ if the cyclisation rate decays strongly enough with polymer length, i.e. with $\nu > \alpha/5$, with $\nu$ the metric exponent (typically 1/2 for random walks and 0.588 for self-avoiding walks) and $\alpha$ the dynamics exponent (typically 1 for Rouse and 2 for reptative dynamics). In these conditions, the fraction of monomers transformed into rings is finite, thus leaving the rest of the monomers available to form a permanently growing linear chain which then drive a runaway reaction. 

We also discover that the runaway transition has deep consequences on the rheology, and triggers an exponential increase for $\kappa < 1$ (or $c/c^* > 0.1$).
Our results suggest that it may be possible to tune the final topological composition of ligated systems by judiciously choosing $c/c^*$. For instance, the most likely regime to form large rings and ring-linear blends~\cite{Kapnistos2008,Halverson2012} is near the transition $c/c^* \simeq 0.1$. Mixing polymer families with different reactive ends further enhances the designability as it introduces different $c^*$ for each family. Our results can be used to optimise the conditions for DNA engineering, e.g., transfection vectors~\cite{Oliynyk2022} ought to be ligated at $c/c^* < 0.1$ whereas synthetic chromosomes assemblies~\cite{Annaluru2014} at large $c/c^*$. Finally, it may be possible to couple dissipative DNA breakage reactions~\cite{Michieletto2022natcomm,DelGrosso2022,Heinen2019} with ATP-consuming ligation to create dense solutions of self-sustained topologically active viscoelastic fluids which would be an interesting active fluid to investigate in the future.  

\begin{acknowledgments}
DM acknowledges the support of the Royal Society via a University Research Fellowship. This project has received support from European Research Council (ERC) under the European Union's Horizon 2020 research and innovation programme (grant agreement No 947918 to DM and No 677532 to ML). The authors acknowledge insightful discussions with Daan Noordermeer and Antonio Valdes who also kindly gifted us with the 1288 plasmid. Source codes are available at \url{https://git.ecdf.ed.ac.uk/taplab/dna-ligation.git}. For the purpose of open access, the author has applied a Creative Commons Attribution (CC BY) licence to any Author Accepted Manuscript version arising from this submission.
\end{acknowledgments}

\appendix

\section{\label{sec:TR}Topology reconstruction algorithm}

We will refer to the Topology Reconstruction code from now on as TR. The code takes as input the instantaneous trajectory and bond list from LAMMPS and checks for newly formed linear and ring chains. The output of the Python code is a file containing the number and length of linear chains that have formed in a given simulation time step. Similar files are produced for the ring chains. These files are then used to calculate the average length and the number of linear/ring chain figures.

The starting point of the TR algorithm is an array $\mathbf b$ of size $N_b \times 2$; each row $b_i = (id_1, id_2)$ represents the IDs of atoms that are bonded within the system:
\[
\mathbf b=
  \begin{bmatrix}
    id_1 & id_2 \\
    id_3 & id_4 \\
    \vdots & \vdots \\
    id_{N_{n-1}} & id_{N_n}
  \end{bmatrix} \, ,
\]

Since not all particles are linked together, some do not appear in the array $\mathbf b$. To avoid operations with large sparse arrays, the matrix $\mathbf b$ is mapped to $\mathbf{ \tilde{b}}$ that contains only indexes from $1$ to the maximum number of atoms connected $M:\mathbf b\to\mathbf{\tilde{b}}$.

The next step of the TR algorithm is to create a connectivity matrix $\mathbf{C}$ based on the list $\mathbf{\tilde{b}}$. Each row of $\mathbf{C}$ represents an atom index and consists of three components $\mathbf{C}(id_{i},:) = (id_{i-1}, id_{i+1}, flag)$. Since in our case a particle can be linked with two more particles the first two components of each row $id_{i-1}, id_{i+1}$ represent the connections of particle $id_{i}$ (note that $id_{i-1}$ and $id_{i+1}$ are not necessarily consecutive in 1D but can be any other particle bonded to particle $i$). The third component, $flag$, takes only the values ${0,1}$ and accounts for the particles $id_i$ that already belong to a polymer chain. The flag column of $\mathbf{C}$ is initialised to zeros. During the reading process of the connectivity matrix, the algorithm switches the flags to 1 of the particles that are already considered to belong in a chain. Rings are extracted in the same manner and a ring is found if the current atom index is the same as the starting atom index. This reading process outputs $N_c$ arrays that have different lengths and each of them contains the particle (mapped) ids that are connected in a polymer chain. The final step of the TR algorithm is to map the atom indexed back to the original ones: $M^{-1}:\mathbf{\tilde{b}}\to\mathbf b$. This algorithm is very generic and can be applied also in cases where the atoms are not initially in polymers as in our case, but rather individual atoms that can connect during the simulation. 

\section{\label{sec:MCstep}Description of the Monte Carlo step}

In this paragraph, we will describe the single Monte Carlo (MC) step, which is repeated a large number of times during the numerical resolution of Eq.~\eqref{eq:smolu} performed using the DSMC algorithm.  With reference to Eq.~\eqref{eq:smolu}, we define $n_f\equiv V \sum_{i=1}^\infty n_i$ (total number of chains). Before the start of the simulation, we give an estimate of the maximum annealing rate $k_\text{max}$ and of the maximum cyclisation rate $k_0^\text{max}$. The exactness of the algorithm does not depend on this initial choice, however, choosing values that are too far from the actual maximum rates can lead to a reduced efficiency~\cite{garcia1987monte}. 

During every MC step, we either attempt to perform a ligation reaction (with probability $p$) or a cyclisation one (with probability $1-p$). The value of $p$ is calculated initially and then updated during the simulation in such a way that the average number densities $n(l)$ satisfy \eqref{eq:smolu}. At the beginning of each MC step, $p$ is evaluated as
\begin{equation}
p^{-1} = 1 + \frac{ 2 N k_o^\text{max}}{(n_f-1) n k_\text{max} } \, .
\label{eq:p}
\end{equation}
We will show below that this choice also guarantees that the simulation samples the correct number of cyclisation and ligation events per unit volume and unit time as required by Eq.~\eqref{eq:smolu}.

We define a \textit{waiting time} variable that is set to zero at the beginning of the simulation. After each reaction, a waiting time increment is generated. These increments are also chosen to guarantee the correct number of ligation and fragmentation reactions per unit of time/volume, as detailed below. We can now describe the MC step, during which the following actions are performed:

 \begin{enumerate}
 
  \item{We evaluate the probability of annealing $p$ according to Eq.~\eqref{eq:p}. The explicit form of $p$, Eq.~\eqref{eq:p}, will be discussed in detail below.}
 
 \item{We pick a random number $0\leq r\leq1$ from a uniform distribution. If $r\leq p$, we attempt a ligation event:}

 \begin{enumerate}
 \item{We pick a pair of elements of the array $\mathbf{m}$, denoted $\alpha,\beta$ at random. Since there are $n_f(n_f-1)$ ordered pairs of chains, the probability of picking a specific pair is $[n_f(n_f-1)]^{-1}$. Let the length associated with these elements be $m_\alpha=i$ and $m_\beta=j$.}
 \item{We evaluate the ligation rate $k_1(i,j)$ for the two chains. If $k_1(i,j)>k_\text{max}$, we set $k_1^\text{max}=k_1(i,j)$ and return to (1). Otherwise, we continue.}
 \item{We pick another random number $r^\prime$, and perform the ligation if $r^\prime \leq k_1(i,j)/k_1^\text{max}$. If ligation is unsuccessful, we return to (1). Otherwise, we continue.}
 \item{We increment the waiting time by $\Delta t^\text{lig}_{i,j} = \frac{2 A N}{ n_f(n_f-1)n k_{ij}}$. Here $A$ is a parameter, the only condition on which is that it must be between $0$ and $1$, as we will discuss in more detail below.}
  \item{After incrementing the waiting time, we update $\mathbf f$ by setting $m_\alpha=0$ and $m_\beta=i+j$.}
 \end{enumerate}
 
 \item{If $r>p$, we attempt a cyclisation event:}
 
  \begin{enumerate}
 \item{We pick a chain $\gamma$ at random with probability $n_f^{-1}$. Let $m_\gamma=l$.}
  \item{We evaluate the cyclisation rate $k_0(l)$. If $k_0(l)>k_0^\text{max}$, set $k_0^\text{max}=k_0(l)$ and return to (1). Otherwise, we continue.}
 \item{We extract another random number $0\leq r^\prime \leq 1$ from a uniform distribution, and perform cyclisation if $r^\prime \leq k_0(l)/k_0^\text{max}$. If cyclisation is unsuccessful, we return to (1). Otherwise, we continue.}
 \item{We increment the waiting time by $\Delta t^\text{cyc}_l = \frac{1-A}{n_f k_o(l)}$, with $A$ defined above in step (2).}
 \item{We record the value of $l$ in $\mathbf r$ and set $m_\gamma=0$. }
 \end{enumerate}
 
 \end{enumerate}

We now prove that the definitions of $p$ (Eq.~\eqref{eq:p}), the waiting time increments $\Delta t^\text{lig}_{i,j}$ (for ligation) and $\Delta t^\text{cyc}_{i,k-i}$ (for cyclisation) give several ligation and cyclisation events per unit time which is consistent with the Smoluchowski equation Eq.~\eqref{eq:smolu}. Over a single MC step, the mean number of ligation events involving the ordered pair of filaments $(\alpha,\beta)$ is
\begin{equation}
\label{eq:nlig1}
 \langle \# L_{\alpha,\beta} \rangle \equiv  \frac p {n_f (n_f-1)} \frac {k_1(m_\alpha,m_\beta)} {k_1^\text{max}} \, .
 \end{equation}
We note that in the algorithm we consider $(m_\alpha,m_\beta)$ as an ordered pair, and thus in \eqref{eq:nlig1} we consider the reaction $(i,j)\to l$ as distinct from $(j,i)\to l$. The mean number of ligation events involving \textit{any two chains} with lengths $i,j$ can be obtained by multiplying the above quantity by $2 (1-\delta_{ij}/2) V^2 n_i n_j$. The factor $2(1-\delta_{ij}/2)$ takes into account the fact that, as mentioned above, for $i\neq j$, there are two ways to perform the ligation, whereas for $i=j$ there is only one. The factor $V^2 n_i n_j$ is the product of the volume fractions of filaments of lengths $i$ and $j$. We thus have
\begin{align}
 \label{eq:nlig2}
&2 V^2 n_i n_j \left(1-\frac{\delta_{ij}} 2\right) \times   \frac p {n_f (n_f-1)} \frac {k_1(i,j)} {k_1^\text{max}} = \notag \\ 
&V k_1(i,j) n_i n_j \left(1-\frac{\delta_{ij}} 2\right) \Delta t \, ,
\end{align}
where we have equated the mean number of ligation events involving any two chains with lengths $i,j$ to the value required by the Smoluchowski equation. Recalling that $n=N/V$, we thus find
 \begin{equation}
 \label{eq:deltat1}
 \Delta t = \frac{2pN}{n_f(n_f-1) n k_1^\text{max}} \, .
 \end{equation}

Eq.~\eqref{eq:deltat1} relates the time interval $\Delta t$ to the probability of ligation. We will now obtain a second equality involving $p$ and $\Delta t$, which will allow us to prove that the expression Eq.~\eqref{eq:p} for $p$ guarantees the correct number of ligation and cyclisation events per unit time.
 
The mean number of cyclisation events involving chains $\gamma$ is
\begin{equation}
\label{eq:ncyc}
\langle \# C_{\gamma} \rangle \equiv \frac{(1-p) k_0(m_\gamma)}{n_f k_0^\text{max}} \, .
\end{equation}
To obtain the mean number of cyclisations of a generic $l-$mer we need to multiply this quantity by $V n_l$, \textit{i.e.}, the volume fraction of filaments of length $l$. Equating this quantity to the expected number of rings formed in a time interval $\Delta t$ we obtain
\begin{equation}
 \label{eq:ncyc1}
V n_l  \times  \frac{(1-p) k_0(l)}{n_f k_0^\text{max}} = k_0(l) n_l V \Delta t \, ,
\end{equation}
and hence
\begin{equation}
\label{eq:deltat2}
\Delta t \equiv \frac{1-p}{n_f k_o^\text{max}} \, .
\end{equation}
By equating the two expressions for $\Delta t$, Eq.~\eqref{eq:deltat1} and Eq.~\eqref{eq:deltat2}, we find Eq.~\eqref{eq:p}. We have thus proven that the latter is the correct expression of $p$, which gives the correct number of cyclisation and ligation events per unit time and unit volume, as required by the Smoluchowski equation.

Finally, we will prove below that the constants $A$ and $1-A$ introduced when calculating the waiting time increments are consistent with Eq.~\eqref{eq:deltat1} and Eq.~\eqref{eq:deltat2}. To show this, it is sufficient to observe that the total time increment during an MC step is:
\begin{align}
\Delta t & = \sum_{0 \leq \alpha < \beta \leq n_f-1}  \langle \# L_{\alpha,\beta} \rangle \Delta t^\text{lig}_{m_\alpha,m_\beta} +  \sum_{i=1}^{m_\gamma-1} \langle \# C_{\gamma} \rangle \Delta t^\text{cyc}_{m_\gamma} \nonumber \\
& =  \sum_{0 \leq \alpha < \beta \leq n_f-1}   \left[\frac {p k_{m_\alpha,m_\beta}} {n_f(n_f-1) k_\text{max}}\right] \left[\frac{2AN }{n_f (n_f-1) n k_{m_\alpha,m_\beta}}\right] \nonumber \\
& \ \ \ + \sum_{i=1}^{m_\gamma-1}  \left[\frac {(1-p)k_o(l)} {n k_o^\text{max}}\right] \frac{1-A}{n_f k_o(l)} \nonumber \\
& = \frac{2 A p N }{n_f (n_f-1) n k_\text{max}} + \frac{(1-A) (1-p)}{n_f k_o^\text{max}}\nonumber \\
\end{align}

One can see that this equality is consistent with Eq.\eqref{eq:deltat1} and Eq.\eqref{eq:deltat2}. We note that the algorithm samples on average the correct kinetics independently of the value of $A$, as long as $0 \leq A \leq 1$. Here we take $A=1$, meaning that the waiting time increment is calculated only after a successful ligation reaction, but not after a successful cyclisation reaction. \\

\section{\label{sec:numIntSmol}Numerical integration of modified Smoluchowski}

Solving the Smoluchowski equation to fit the data from MD simulations consists of two main parts: 
\begin{enumerate}
  \item{We create an objective function for the \emph{lsqcurvefit} (called \emph{Obj\_smoluchowski}) that takes as input the array of initial coefficient guess $K_0 =(\kappa_1, \kappa_0)$ and the time data array $xdata$. It returns the average length as a function of time, array $ydata$. In the objective function:}
  \begin{enumerate}
      \item{An array $\mathbf L= \mathbf{n} \cdot l_0$ is initialised where $n = \{1,2,\dots,N_c=200\}$ and $l_0=174$. This represents the set of lengths that can be found in the system (recall that we initialise our MD simulations with 200 chains of 174 beads each). Also, the arrays with the number density of linear and ring chains are initialised as follows,
      $\mathbf{n_{l_0}} = (N_c/vol,0,\dots,0)_{1\times N_c}
      $} and $\mathbf{n_{r_0}} = (0,\dots,0)_{1\times N_c}$
      since initially all the molecules are linear chains.  Here, $vol$ denotes the volume of the simulation box.
      \item{\emph{for} $t$ = \{1 to simulation final step time\} do
      \emph{call} $ (\mathbf {n_{L_{new}}},\mathbf {n_{R_{new}}})=exEuler\_smoluchowski \left(\mathbf {n_L},\mathbf {n_R},K\right)$ function (see point 2 below)}
      \item{update arrays $\mathbf {n_L}=\mathbf {n_{L_{new}}}$ and $\mathbf {n_R}=\mathbf {n_{R_{new}}}$. Calculate the total average length $\mathbf{l_{total}}$ as
      $$
      \mathbf {l_{total}(t)} = \frac{\mathbf {n_{L_{new}}} \cdot \mathbf L+ \mathbf {n_{R_{new}}} \cdot \mathbf L}{\sum_i^{N_m}{n_{L_{new}}^i}+ \sum_i^{N_m}{n_{R_{new}}^i}}
      $$}
      \item{\emph{exit for loop} and parse $\mathbf {l_{total}(t)}$ to $ydata$}
  \end{enumerate}
 \item{The exEuler\_smoluchowski function takes as input the initial number densities of linear and ring chains and the reaction rates $\mathbf {n_L},\mathbf {n_R}, K=(\kappa_1,\kappa_0)$. Based on the given rates K, it outputs the final number density arrays $\mathbf {n_{L_{new}}},\mathbf {n_{R_{new}}}$, after the reactions have taken place.  When this function is called, the number density of linear and ring chains of each population are updated according to Eq.~\eqref{eq:smolu}. The monomer, dimer, and so on populations are increased according to the first two terms of Eq.~\eqref{eq:smolu} while the number of them that is converted into rings is subtracted by the $\mathbf {n_{L_{new}}}$ and added to the $\mathbf {n_{R_{new}}}$ array. \\
 In the first two terms of Eq.~\eqref{eq:smolu1} the rate $k_1(i,j)$ is not a scalar quantity by rather a matrix that follows the relation Eq.~\eqref{eq:smolu_rates}. The extracted coefficient against which the fitting is optimised is the scalar $\kappa_1$. Similarly, for the sink term of Eq.~\eqref{eq:smolu1}-~\eqref{eq:smolu2}, the equation $k_0(l) = \kappa_0l^{-4\nu}$ is used and the fitting coefficient exported is the scalar $\kappa_0$.
 }
\end{enumerate}
The coefficients $K$ are updated iteratively by the \emph{lsqcurvefit} algorithm to best fit the data. Once the optimum values are obtained the algorithm terminates. 

\bibliographystyle{apsrev4-2}

\begin{thebibliography}{59}%
\makeatletter
\providecommand \@ifxundefined [1]{%
 \@ifx{#1\undefined}
}%
\providecommand \@ifnum [1]{%
 \ifnum #1\expandafter \@firstoftwo
 \else \expandafter \@secondoftwo
 \fi
}%
\providecommand \@ifx [1]{%
 \ifx #1\expandafter \@firstoftwo
 \else \expandafter \@secondoftwo
 \fi
}%
\providecommand \natexlab [1]{#1}%
\providecommand \enquote  [1]{``#1''}%
\providecommand \bibnamefont  [1]{#1}%
\providecommand \bibfnamefont [1]{#1}%
\providecommand \citenamefont [1]{#1}%
\providecommand \href@noop [0]{\@secondoftwo}%
\providecommand \href [0]{\begingroup \@sanitize@url \@href}%
\providecommand \@href[1]{\@@startlink{#1}\@@href}%
\providecommand \@@href[1]{\endgroup#1\@@endlink}%
\providecommand \@sanitize@url [0]{\catcode `\\12\catcode `\$12\catcode
  `\&12\catcode `\#12\catcode `\^12\catcode `\_12\catcode `\%12\relax}%
\providecommand \@@startlink[1]{}%
\providecommand \@@endlink[0]{}%
\providecommand \url  [0]{\begingroup\@sanitize@url \@url }%
\providecommand \@url [1]{\endgroup\@href {#1}{\urlprefix }}%
\providecommand \urlprefix  [0]{URL }%
\providecommand \Eprint [0]{\href }%
\providecommand \doibase [0]{https://doi.org/}%
\providecommand \selectlanguage [0]{\@gobble}%
\providecommand \bibinfo  [0]{\@secondoftwo}%
\providecommand \bibfield  [0]{\@secondoftwo}%
\providecommand \translation [1]{[#1]}%
\providecommand \BibitemOpen [0]{}%
\providecommand \bibitemStop [0]{}%
\providecommand \bibitemNoStop [0]{.\EOS\space}%
\providecommand \EOS [0]{\spacefactor3000\relax}%
\providecommand \BibitemShut  [1]{\csname bibitem#1\endcsname}%
\let\auto@bib@innerbib\@empty
\bibitem [{\citenamefont {Rubinstein}\ and\ \citenamefont
  {Colby}(2003)}]{Rubinsteina}%
  \BibitemOpen
  \bibfield  {author} {\bibinfo {author} {\bibfnamefont {M.}~\bibnamefont
  {Rubinstein}}\ and\ \bibinfo {author} {\bibfnamefont {H.~R.}\ \bibnamefont
  {Colby}},\ }\href@noop {} {\emph {\bibinfo {title} {{Polymer Physics}}}}\
  (\bibinfo  {publisher} {Oxford University Press},\ \bibinfo {year}
  {2003})\BibitemShut {NoStop}%
\bibitem [{\citenamefont {Oliynyk}\ and\ \citenamefont
  {Church}(2022)}]{Oliynyk2022}%
  \BibitemOpen
  \bibfield  {author} {\bibinfo {author} {\bibfnamefont {R.~T.}\ \bibnamefont
  {Oliynyk}}\ and\ \bibinfo {author} {\bibfnamefont {G.~M.}\ \bibnamefont
  {Church}},\ }\href@noop {} {\bibfield  {journal} {\bibinfo  {journal}
  {Communications biology}\ }\textbf {\bibinfo {volume} {5}} (\bibinfo {year}
  {2022})}\BibitemShut {NoStop}%
\bibitem [{\citenamefont {Alberts}\ \emph {et~al.}(2014)\citenamefont
  {Alberts}, \citenamefont {Johnson}, \citenamefont {Lewis}, \citenamefont
  {Morgan},\ and\ \citenamefont {Raff}}]{Alberts2014}%
  \BibitemOpen
  \bibfield  {author} {\bibinfo {author} {\bibfnamefont {B.}~\bibnamefont
  {Alberts}}, \bibinfo {author} {\bibfnamefont {A.}~\bibnamefont {Johnson}},
  \bibinfo {author} {\bibfnamefont {J.}~\bibnamefont {Lewis}}, \bibinfo
  {author} {\bibfnamefont {D.}~\bibnamefont {Morgan}},\ and\ \bibinfo {author}
  {\bibfnamefont {M.}~\bibnamefont {Raff}},\ }\href
  {http://books.google.com/books?id=1ZUDoQEACAAJ{\&}pgis=1} {\emph {\bibinfo
  {title} {{Molecular Biology of the Cell}}}}\ (\bibinfo  {publisher} {Taylor
  {\&} Francis},\ \bibinfo {year} {2014})\ p.\ \bibinfo {pages}
  {1464}\BibitemShut {NoStop}%
\bibitem [{\citenamefont {Flory}(1936)}]{Flory1936}%
  \BibitemOpen
  \bibfield  {author} {\bibinfo {author} {\bibfnamefont {P.~J.}\ \bibnamefont
  {Flory}},\ }\href@noop {} {\bibfield  {journal} {\bibinfo  {journal} {Journal
  of the American Chemical Society}\ }\textbf {\bibinfo {volume} {58}},\
  \bibinfo {pages} {1877} (\bibinfo {year} {1936})}\BibitemShut {NoStop}%
\bibitem [{\citenamefont {Cates}\ and\ \citenamefont
  {Candau}(2001)}]{Cates2001}%
  \BibitemOpen
  \bibfield  {author} {\bibinfo {author} {\bibfnamefont {M.~E.}\ \bibnamefont
  {Cates}}\ and\ \bibinfo {author} {\bibfnamefont {S.~J.}\ \bibnamefont
  {Candau}},\ }\href@noop {} {\bibfield  {journal} {\bibinfo  {journal} {EPL}\
  }\textbf {\bibinfo {volume} {55}},\ \bibinfo {pages} {887} (\bibinfo {year}
  {2001})}\BibitemShut {NoStop}%
\bibitem [{\citenamefont {{Vafabakhsh Reza, Ha}}(2012)}]{VafabakhshRezaHa2012}%
  \BibitemOpen
  \bibfield  {author} {\bibinfo {author} {\bibfnamefont {T.}~\bibnamefont
  {{Vafabakhsh Reza, Ha}}},\ }\href@noop {} {\bibfield  {journal} {\bibinfo
  {journal} {Science}\ }\textbf {\bibinfo {volume} {337}},\ \bibinfo {pages}
  {1097} (\bibinfo {year} {2012})}\BibitemShut {NoStop}%
\bibitem [{\citenamefont {Zhou}\ \emph {et~al.}(2012)\citenamefont {Zhou},
  \citenamefont {Woo}, \citenamefont {Cok}, \citenamefont {Wang}, \citenamefont
  {Olsen},\ and\ \citenamefont {Johnson}}]{Zhou2012}%
  \BibitemOpen
  \bibfield  {author} {\bibinfo {author} {\bibfnamefont {H.}~\bibnamefont
  {Zhou}}, \bibinfo {author} {\bibfnamefont {J.}~\bibnamefont {Woo}}, \bibinfo
  {author} {\bibfnamefont {A.~M.}\ \bibnamefont {Cok}}, \bibinfo {author}
  {\bibfnamefont {M.}~\bibnamefont {Wang}}, \bibinfo {author} {\bibfnamefont
  {B.~D.}\ \bibnamefont {Olsen}},\ and\ \bibinfo {author} {\bibfnamefont
  {J.~A.}\ \bibnamefont {Johnson}},\ }\href@noop {} {\bibfield  {journal}
  {\bibinfo  {journal} {Proc. Natl. Acad. Sci USA}\ }\textbf {\bibinfo {volume}
  {109}},\ \bibinfo {pages} {19119} (\bibinfo {year} {2012})}\BibitemShut
  {NoStop}%
\bibitem [{\citenamefont {Jacobson}\ and\ \citenamefont
  {Stockmayer}(1950)}]{Jacobson1950}%
  \BibitemOpen
  \bibfield  {author} {\bibinfo {author} {\bibfnamefont {H.}~\bibnamefont
  {Jacobson}}\ and\ \bibinfo {author} {\bibfnamefont {W.~H.}\ \bibnamefont
  {Stockmayer}},\ }\href@noop {} {\bibfield  {journal} {\bibinfo  {journal}
  {The Journal of Chemical Physics}\ }\textbf {\bibinfo {volume} {18}},\
  \bibinfo {pages} {1600 } (\bibinfo {year} {1950})}\BibitemShut {NoStop}%
\bibitem [{\citenamefont {Flory}\ and\ \citenamefont
  {Semlyen}(1966)}]{Flory1966}%
  \BibitemOpen
  \bibfield  {author} {\bibinfo {author} {\bibfnamefont {P.~J.}\ \bibnamefont
  {Flory}}\ and\ \bibinfo {author} {\bibfnamefont {J.~A.}\ \bibnamefont
  {Semlyen}},\ }\href@noop {} {\bibfield  {journal} {\bibinfo  {journal}
  {Journal of the American Chemical Society}\ }\textbf {\bibinfo {volume}
  {88}},\ \bibinfo {pages} {3209} (\bibinfo {year} {1966})}\BibitemShut
  {NoStop}%
\bibitem [{\citenamefont {Suematsu}\ and\ \citenamefont
  {Okamoto}(1992)}]{Suematsu1992}%
  \BibitemOpen
  \bibfield  {author} {\bibinfo {author} {\bibfnamefont {K.}~\bibnamefont
  {Suematsu}}\ and\ \bibinfo {author} {\bibfnamefont {T.}~\bibnamefont
  {Okamoto}},\ }\href {https://doi.org/10.1007/BF00665984} {\bibfield
  {journal} {\bibinfo  {journal} {Colloid {\&} Polymer Science}\ }\textbf
  {\bibinfo {volume} {270}},\ \bibinfo {pages} {421} (\bibinfo {year}
  {1992})}\BibitemShut {NoStop}%
\bibitem [{\citenamefont {Chen}\ and\ \citenamefont
  {Dormidontova}(2004)}]{Dormidontova2004}%
  \BibitemOpen
  \bibfield  {author} {\bibinfo {author} {\bibfnamefont {C.~C.}\ \bibnamefont
  {Chen}}\ and\ \bibinfo {author} {\bibfnamefont {E.~E.}\ \bibnamefont
  {Dormidontova}},\ }\href {https://doi.org/10.1021/ma035405t} {\bibfield
  {journal} {\bibinfo  {journal} {Macromolecules}\ }\textbf {\bibinfo {volume}
  {37}},\ \bibinfo {pages} {3905} (\bibinfo {year} {2004})}\BibitemShut
  {NoStop}%
\bibitem [{\citenamefont {Ercolani}\ and\ \citenamefont
  {Stefano}(2008)}]{Ercolani2008}%
  \BibitemOpen
  \bibfield  {author} {\bibinfo {author} {\bibfnamefont {G.}~\bibnamefont
  {Ercolani}}\ and\ \bibinfo {author} {\bibfnamefont {D.}~\bibnamefont
  {Stefano}},\ }\href {https://doi.org/10.1021/jp711389t} {\bibfield  {journal}
  {\bibinfo  {journal} {Journal of Physical Chemistry B}\ }\textbf {\bibinfo
  {volume} {112}},\ \bibinfo {pages} {4662} (\bibinfo {year}
  {2008})}\BibitemShut {NoStop}%
\bibitem [{\citenamefont {Madeleine-Perdrillat}\ \emph
  {et~al.}(2014)\citenamefont {Madeleine-Perdrillat}, \citenamefont
  {Delor-Jestin},\ and\ \citenamefont {{De Sainte
  Claire}}}]{Madeleine-Perdrillat2014a}%
  \BibitemOpen
  \bibfield  {author} {\bibinfo {author} {\bibfnamefont {C.}~\bibnamefont
  {Madeleine-Perdrillat}}, \bibinfo {author} {\bibfnamefont {F.}~\bibnamefont
  {Delor-Jestin}},\ and\ \bibinfo {author} {\bibfnamefont {P.}~\bibnamefont
  {{De Sainte Claire}}},\ }\href@noop {} {\bibfield  {journal} {\bibinfo
  {journal} {Journal of Physical Chemistry B}\ }\textbf {\bibinfo {volume}
  {118}},\ \bibinfo {pages} {330} (\bibinfo {year} {2014})}\BibitemShut
  {NoStop}%
\bibitem [{\citenamefont {{Di Stefano}}\ and\ \citenamefont
  {Mandolini}(2019)}]{DiStefano2019}%
  \BibitemOpen
  \bibfield  {author} {\bibinfo {author} {\bibfnamefont {S.}~\bibnamefont {{Di
  Stefano}}}\ and\ \bibinfo {author} {\bibfnamefont {L.}~\bibnamefont
  {Mandolini}},\ }\href@noop {} {\bibfield  {journal} {\bibinfo  {journal}
  {Physical Chemistry Chemical Physics}\ }\textbf {\bibinfo {volume} {21}},\
  \bibinfo {pages} {955} (\bibinfo {year} {2019})}\BibitemShut {NoStop}%
\bibitem [{\citenamefont {Kricheldorf}\ \emph {et~al.}(2020)\citenamefont
  {Kricheldorf}, \citenamefont {Weidner},\ and\ \citenamefont
  {Scheliga}}]{Kricheldorf2020}%
  \BibitemOpen
  \bibfield  {author} {\bibinfo {author} {\bibfnamefont {H.~R.}\ \bibnamefont
  {Kricheldorf}}, \bibinfo {author} {\bibfnamefont {S.~M.}\ \bibnamefont
  {Weidner}},\ and\ \bibinfo {author} {\bibfnamefont {F.}~\bibnamefont
  {Scheliga}},\ }\href@noop {} {\bibfield  {journal} {\bibinfo  {journal}
  {Polymer Chemistry}\ }\textbf {\bibinfo {volume} {11}},\ \bibinfo {pages}
  {2595} (\bibinfo {year} {2020})}\BibitemShut {NoStop}%
\bibitem [{\citenamefont {Lang}\ and\ \citenamefont {Kumar}(2021)}]{Lang2021}%
  \BibitemOpen
  \bibfield  {author} {\bibinfo {author} {\bibfnamefont {M.}~\bibnamefont
  {Lang}}\ and\ \bibinfo {author} {\bibfnamefont {K.~S.}\ \bibnamefont
  {Kumar}},\ }\href@noop {} {\bibfield  {journal} {\bibinfo  {journal}
  {Macromolecules}\ }\textbf {\bibinfo {volume} {54}},\ \bibinfo {pages} {7021}
  (\bibinfo {year} {2021})}\BibitemShut {NoStop}%
\bibitem [{\citenamefont {Kricheldorf}\ \emph {et~al.}(2022)\citenamefont
  {Kricheldorf}, \citenamefont {Weidner},\ and\ \citenamefont
  {Falkenhagen}}]{Kricheldorf2022}%
  \BibitemOpen
  \bibfield  {author} {\bibinfo {author} {\bibfnamefont {H.~R.}\ \bibnamefont
  {Kricheldorf}}, \bibinfo {author} {\bibfnamefont {S.~M.}\ \bibnamefont
  {Weidner}},\ and\ \bibinfo {author} {\bibfnamefont {J.}~\bibnamefont
  {Falkenhagen}},\ }\href {https://doi.org/10.1039/d1py01679b} {\bibfield
  {journal} {\bibinfo  {journal} {Polymer Chemistry}\ }\textbf {\bibinfo
  {volume} {13}},\ \bibinfo {pages} {1177} (\bibinfo {year}
  {2022})}\BibitemShut {NoStop}%
\bibitem [{\citenamefont {Smoluchowski}(1918)}]{smoluchowski1918versuch}%
  \BibitemOpen
  \bibfield  {author} {\bibinfo {author} {\bibfnamefont {M.~v.}\ \bibnamefont
  {Smoluchowski}},\ }\href@noop {} {\bibfield  {journal} {\bibinfo  {journal}
  {Zeitschrift f{\"u}r physikalische Chemie}\ }\textbf {\bibinfo {volume}
  {92}},\ \bibinfo {pages} {129} (\bibinfo {year} {1918})}\BibitemShut
  {NoStop}%
\bibitem [{\citenamefont {Ziff}(1980)}]{ziff1980kinetics}%
  \BibitemOpen
  \bibfield  {author} {\bibinfo {author} {\bibfnamefont {R.~M.}\ \bibnamefont
  {Ziff}},\ }\href@noop {} {\bibfield  {journal} {\bibinfo  {journal} {Journal
  of Statistical Physics}\ }\textbf {\bibinfo {volume} {23}},\ \bibinfo {pages}
  {241} (\bibinfo {year} {1980})}\BibitemShut {NoStop}%
\bibitem [{\citenamefont {Shimada}\ and\ \citenamefont
  {Yamakawa}(1984)}]{Shimada1984}%
  \BibitemOpen
  \bibfield  {author} {\bibinfo {author} {\bibfnamefont {J.}~\bibnamefont
  {Shimada}}\ and\ \bibinfo {author} {\bibfnamefont {H.}~\bibnamefont
  {Yamakawa}},\ }\href@noop {} {\bibfield  {journal} {\bibinfo  {journal}
  {Macromolecules}\ }\textbf {\bibinfo {volume} {17}},\ \bibinfo {pages} {689}
  (\bibinfo {year} {1984})}\BibitemShut {NoStop}%
\bibitem [{\citenamefont {Kremer}\ and\ \citenamefont
  {Grest}(1990)}]{Kremer1990}%
  \BibitemOpen
  \bibfield  {author} {\bibinfo {author} {\bibfnamefont {K.}~\bibnamefont
  {Kremer}}\ and\ \bibinfo {author} {\bibfnamefont {G.~S.}\ \bibnamefont
  {Grest}},\ }\href@noop {} {\bibfield  {journal} {\bibinfo  {journal} {The
  Journal of Chemical Physics}\ }\textbf {\bibinfo {volume} {92}},\ \bibinfo
  {pages} {5057} (\bibinfo {year} {1990})}\BibitemShut {NoStop}%
\bibitem [{\citenamefont {Plimpton}(1995)}]{Plimpton1995}%
  \BibitemOpen
  \bibfield  {author} {\bibinfo {author} {\bibfnamefont {S.}~\bibnamefont
  {Plimpton}},\ }\href@noop {} {\bibfield  {journal} {\bibinfo  {journal} {J.
  Comp. Phys.}\ }\textbf {\bibinfo {volume} {117}},\ \bibinfo {pages} {1}
  (\bibinfo {year} {1995})}\BibitemShut {NoStop}%
\bibitem [{\citenamefont {Garcia}\ \emph {et~al.}(1987)\citenamefont {Garcia},
  \citenamefont {Van Den~Broeck}, \citenamefont {Aertsens},\ and\ \citenamefont
  {Serneels}}]{garcia1987monte}%
  \BibitemOpen
  \bibfield  {author} {\bibinfo {author} {\bibfnamefont {A.~L.}\ \bibnamefont
  {Garcia}}, \bibinfo {author} {\bibfnamefont {C.}~\bibnamefont {Van
  Den~Broeck}}, \bibinfo {author} {\bibfnamefont {M.}~\bibnamefont
  {Aertsens}},\ and\ \bibinfo {author} {\bibfnamefont {R.}~\bibnamefont
  {Serneels}},\ }\href@noop {} {\bibfield  {journal} {\bibinfo  {journal}
  {Physica A}\ }\textbf {\bibinfo {volume} {143}},\ \bibinfo {pages} {535}
  (\bibinfo {year} {1987})}\BibitemShut {NoStop}%
\bibitem [{\citenamefont {Liffman}(1992)}]{liffman1992direct}%
  \BibitemOpen
  \bibfield  {author} {\bibinfo {author} {\bibfnamefont {K.}~\bibnamefont
  {Liffman}},\ }\href@noop {} {\bibfield  {journal} {\bibinfo  {journal} {J.
  Comput. Phys.}\ }\textbf {\bibinfo {volume} {100}},\ \bibinfo {pages} {116}
  (\bibinfo {year} {1992})}\BibitemShut {NoStop}%
\bibitem [{\citenamefont {Kruis}\ \emph {et~al.}(2000)\citenamefont {Kruis},
  \citenamefont {Maisels},\ and\ \citenamefont {Fissan}}]{kruis2000direct}%
  \BibitemOpen
  \bibfield  {author} {\bibinfo {author} {\bibfnamefont {F.~E.}\ \bibnamefont
  {Kruis}}, \bibinfo {author} {\bibfnamefont {A.}~\bibnamefont {Maisels}},\
  and\ \bibinfo {author} {\bibfnamefont {H.}~\bibnamefont {Fissan}},\
  }\href@noop {} {\bibfield  {journal} {\bibinfo  {journal} {AIChE J.}\
  }\textbf {\bibinfo {volume} {46}},\ \bibinfo {pages} {1735} (\bibinfo {year}
  {2000})}\BibitemShut {NoStop}%
\bibitem [{\citenamefont {Tran}\ \emph {et~al.}(2023)\citenamefont {Tran},
  \citenamefont {Sorichetti}, \citenamefont {Pehau-Arnaudet}, \citenamefont
  {Lenz},\ and\ \citenamefont {Leduc}}]{tran2023fragmentation}%
  \BibitemOpen
  \bibfield  {author} {\bibinfo {author} {\bibfnamefont {Q.~D.}\ \bibnamefont
  {Tran}}, \bibinfo {author} {\bibfnamefont {V.}~\bibnamefont {Sorichetti}},
  \bibinfo {author} {\bibfnamefont {G.}~\bibnamefont {Pehau-Arnaudet}},
  \bibinfo {author} {\bibfnamefont {M.}~\bibnamefont {Lenz}},\ and\ \bibinfo
  {author} {\bibfnamefont {C.}~\bibnamefont {Leduc}},\ }\href@noop {}
  {\bibfield  {journal} {\bibinfo  {journal} {Physical Review X}\ }\textbf
  {\bibinfo {volume} {13}},\ \bibinfo {pages} {011014} (\bibinfo {year}
  {2023})}\BibitemShut {NoStop}%
\bibitem [{\citenamefont {Robertson}\ \emph {et~al.}(2006)\citenamefont
  {Robertson}, \citenamefont {Laib},\ and\ \citenamefont
  {Smith}}]{Robertson2006}%
  \BibitemOpen
  \bibfield  {author} {\bibinfo {author} {\bibfnamefont {R.~M.}\ \bibnamefont
  {Robertson}}, \bibinfo {author} {\bibfnamefont {S.}~\bibnamefont {Laib}},\
  and\ \bibinfo {author} {\bibfnamefont {D.~E.}\ \bibnamefont {Smith}},\ }\href
  {http://www.pubmedcentral.nih.gov/articlerender.fcgi?artid=1450111{\&}tool=pmcentrez{\&}rendertype=abstract}
  {\bibfield  {journal} {\bibinfo  {journal} {Proc. Natl. Acad. Sci. USA}\
  }\textbf {\bibinfo {volume} {103}},\ \bibinfo {pages} {7310} (\bibinfo {year}
  {2006})}\BibitemShut {NoStop}%
\bibitem [{\citenamefont {Taylor}\ and\ \citenamefont
  {Hagerman}(1990)}]{Taylor1990}%
  \BibitemOpen
  \bibfield  {author} {\bibinfo {author} {\bibfnamefont {W.~H.}\ \bibnamefont
  {Taylor}}\ and\ \bibinfo {author} {\bibfnamefont {P.~J.}\ \bibnamefont
  {Hagerman}},\ }\href@noop {} {\bibfield  {journal} {\bibinfo  {journal}
  {Journal of Molecular Biology}\ }\textbf {\bibinfo {volume} {212}},\ \bibinfo
  {pages} {363} (\bibinfo {year} {1990})}\BibitemShut {NoStop}%
\bibitem [{\citenamefont {Bates}\ and\ \citenamefont
  {Maxwell}(2005)}]{Bates2005}%
  \BibitemOpen
  \bibfield  {author} {\bibinfo {author} {\bibfnamefont {A.}~\bibnamefont
  {Bates}}\ and\ \bibinfo {author} {\bibfnamefont {A.}~\bibnamefont
  {Maxwell}},\ }\href
  {http://books.google.com/books?hl=en{\&}lr={\&}id=WGBAGyzvQOUC{\&}oi=fnd{\&}pg=PR17{\&}dq=DNA+topology{\&}ots=TUaM8kASav{\&}sig=56tWxeOcV-zwI3l9c030FtWj1Y0}
  {\emph {\bibinfo {title} {{DNA topology}}}}\ (\bibinfo  {publisher} {Oxford
  University Press},\ \bibinfo {year} {2005})\BibitemShut {NoStop}%
\bibitem [{\citenamefont {Crocker}\ \emph {et~al.}(2000)\citenamefont
  {Crocker}, \citenamefont {Valentine}, \citenamefont {Weeks}, \citenamefont
  {Gisler}, \citenamefont {Kaplan}, \citenamefont {Yodh},\ and\ \citenamefont
  {Weitz}}]{Crocker2000}%
  \BibitemOpen
  \bibfield  {author} {\bibinfo {author} {\bibfnamefont {J.~C.}\ \bibnamefont
  {Crocker}}, \bibinfo {author} {\bibfnamefont {M.~T.}\ \bibnamefont
  {Valentine}}, \bibinfo {author} {\bibfnamefont {E.~R.}\ \bibnamefont
  {Weeks}}, \bibinfo {author} {\bibfnamefont {T.}~\bibnamefont {Gisler}},
  \bibinfo {author} {\bibfnamefont {P.~D.}\ \bibnamefont {Kaplan}}, \bibinfo
  {author} {\bibfnamefont {A.~G.}\ \bibnamefont {Yodh}},\ and\ \bibinfo
  {author} {\bibfnamefont {D.~A.}\ \bibnamefont {Weitz}},\ }\href@noop {}
  {\bibfield  {journal} {\bibinfo  {journal} {Phys. Rev. Lett.}\ }\textbf
  {\bibinfo {volume} {85}},\ \bibinfo {pages} {888} (\bibinfo {year}
  {2000})}\BibitemShut {NoStop}%
\bibitem [{\citenamefont {Hansen}\ and\ \citenamefont
  {McDonald}(2013)}]{hansen2013theory}%
  \BibitemOpen
  \bibfield  {author} {\bibinfo {author} {\bibfnamefont {J.-P.}\ \bibnamefont
  {Hansen}}\ and\ \bibinfo {author} {\bibfnamefont {I.~R.}\ \bibnamefont
  {McDonald}},\ }\href@noop {} {\emph {\bibinfo {title} {Theory of simple
  liquids: with applications to soft matter}}}\ (\bibinfo  {publisher}
  {Academic press},\ \bibinfo {year} {2013})\BibitemShut {NoStop}%
\bibitem [{\citenamefont {{De Gennes}}(1982{\natexlab{a}})}]{DeGennes1982a}%
  \BibitemOpen
  \bibfield  {author} {\bibinfo {author} {\bibfnamefont {P.~G.}\ \bibnamefont
  {{De Gennes}}},\ }\href@noop {} {\bibfield  {journal} {\bibinfo  {journal}
  {The Journal of Chemical Physics}\ }\textbf {\bibinfo {volume} {76}},\
  \bibinfo {pages} {3316} (\bibinfo {year} {1982}{\natexlab{a}})}\BibitemShut
  {NoStop}%
\bibitem [{\citenamefont {Grosberg}\ \emph {et~al.}(1982)\citenamefont
  {Grosberg}, \citenamefont {Khalatur},\ and\ \citenamefont
  {Khokhlov}}]{grosberg1982polymeric}%
  \BibitemOpen
  \bibfield  {author} {\bibinfo {author} {\bibfnamefont {A.~Y.}\ \bibnamefont
  {Grosberg}}, \bibinfo {author} {\bibfnamefont {P.~G.}\ \bibnamefont
  {Khalatur}},\ and\ \bibinfo {author} {\bibfnamefont {A.~R.}\ \bibnamefont
  {Khokhlov}},\ }\href@noop {} {\bibfield  {journal} {\bibinfo  {journal} {Die
  Makromolekulare Chemie, Rapid Communications}\ }\textbf {\bibinfo {volume}
  {3}},\ \bibinfo {pages} {709} (\bibinfo {year} {1982})}\BibitemShut {NoStop}%
\bibitem [{\citenamefont {Doi}\ and\ \citenamefont {Edwards}(1988)}]{Doi1988}%
  \BibitemOpen
  \bibfield  {author} {\bibinfo {author} {\bibfnamefont {M.}~\bibnamefont
  {Doi}}\ and\ \bibinfo {author} {\bibfnamefont {S.}~\bibnamefont {Edwards}},\
  }\href@noop {} {\emph {\bibinfo {title} {{The theory of polymer dynamics}}}}\
  (\bibinfo  {publisher} {Oxford University Press},\ \bibinfo {year}
  {1988})\BibitemShut {NoStop}%
\bibitem [{\citenamefont {Rosa}\ \emph {et~al.}(2010)\citenamefont {Rosa},
  \citenamefont {Becker},\ and\ \citenamefont {Everaers}}]{Rosa2010}%
  \BibitemOpen
  \bibfield  {author} {\bibinfo {author} {\bibfnamefont {A.}~\bibnamefont
  {Rosa}}, \bibinfo {author} {\bibfnamefont {N.~B.}\ \bibnamefont {Becker}},\
  and\ \bibinfo {author} {\bibfnamefont {R.}~\bibnamefont {Everaers}},\ }\href
  {http://dx.@doi.org/10.1016/j.bpj.2010.01.054} {\bibfield  {journal}
  {\bibinfo  {journal} {Biophys. J.}\ }\textbf {\bibinfo {volume} {98}},\
  \bibinfo {pages} {2410} (\bibinfo {year} {2010})}\BibitemShut {NoStop}%
\bibitem [{\citenamefont {Rosa}\ and\ \citenamefont
  {Everaers}(2008)}]{Rosa2008}%
  \BibitemOpen
  \bibfield  {author} {\bibinfo {author} {\bibfnamefont {A.}~\bibnamefont
  {Rosa}}\ and\ \bibinfo {author} {\bibfnamefont {R.}~\bibnamefont
  {Everaers}},\ }\href
  {http://www.pubmedcentral.nih.gov/articlerender.fcgi?artid=2515109{\&}tool=pmcentrez{\&}rendertype=abstract}
  {\bibfield  {journal} {\bibinfo  {journal} {PLoS computational biology}\
  }\textbf {\bibinfo {volume} {4}},\ \bibinfo {pages} {1} (\bibinfo {year}
  {2008})}\BibitemShut {NoStop}%
\bibitem [{\citenamefont {\textbf{V. Sorichetti}}\ and\ \citenamefont
  {Lenz}(2023)}]{sorichetti2023transverse}%
  \BibitemOpen
  \bibfield  {author} {\bibinfo {author} {\bibnamefont {\textbf{V.
  Sorichetti}}}\ and\ \bibinfo {author} {\bibfnamefont {M.}~\bibnamefont
  {Lenz}},\ }\href {https://doi.org/10.1103/PhysRevLett.131.228401} {\bibfield
  {journal} {\bibinfo  {journal} {Phys. Rev. Lett.}\ }\textbf {\bibinfo
  {volume} {131}},\ \bibinfo {pages} {228401} (\bibinfo {year}
  {2023})}\BibitemShut {NoStop}%
\bibitem [{\citenamefont {van Dongen}\ and\ \citenamefont
  {Ernst}(1984)}]{vandongen1984kinetics}%
  \BibitemOpen
  \bibfield  {author} {\bibinfo {author} {\bibfnamefont {P.~G.~J.}\
  \bibnamefont {van Dongen}}\ and\ \bibinfo {author} {\bibfnamefont {M.~H.}\
  \bibnamefont {Ernst}},\ }\href {https://doi.org/10.1007/BF01011836}
  {\bibfield  {journal} {\bibinfo  {journal} {J. Stat. Phys.}\ }\textbf
  {\bibinfo {volume} {37}},\ \bibinfo {pages} {301} (\bibinfo {year}
  {1984})}\BibitemShut {NoStop}%
\bibitem [{\citenamefont {Van~Dongen}\ and\ \citenamefont
  {Ernst}(1985)}]{vandongen1985dynamic}%
  \BibitemOpen
  \bibfield  {author} {\bibinfo {author} {\bibfnamefont {P.}~\bibnamefont
  {Van~Dongen}}\ and\ \bibinfo {author} {\bibfnamefont {M.}~\bibnamefont
  {Ernst}},\ }\href@noop {} {\bibfield  {journal} {\bibinfo  {journal}
  {Physical review letters}\ }\textbf {\bibinfo {volume} {54}},\ \bibinfo
  {pages} {1396} (\bibinfo {year} {1985})}\BibitemShut {NoStop}%
\bibitem [{\citenamefont {@doi}\ and\ \citenamefont
  {Edwards}(1988)}]{Doi1988b}%
  \BibitemOpen
  \bibfield  {author} {\bibinfo {author} {\bibfnamefont {M.}~\bibnamefont
  {@doi}}\ and\ \bibinfo {author} {\bibfnamefont {S.}~\bibnamefont {Edwards}},\
  }\href@noop {} {\bibinfo {title} {{The Theory of Polymer Dynamics}}}
  (\bibinfo {year} {1988})\BibitemShut {NoStop}%
\bibitem [{\citenamefont {{De Gennes}}(1982{\natexlab{b}})}]{DeGennes1982}%
  \BibitemOpen
  \bibfield  {author} {\bibinfo {author} {\bibfnamefont {P.~G.}\ \bibnamefont
  {{De Gennes}}},\ }\href@noop {} {\bibfield  {journal} {\bibinfo  {journal}
  {The Journal of Chemical Physics}\ }\textbf {\bibinfo {volume} {76}},\
  \bibinfo {pages} {3322} (\bibinfo {year} {1982}{\natexlab{b}})}\BibitemShut
  {NoStop}%
\bibitem [{\citenamefont {Meakin}\ and\ \citenamefont
  {Ernst}(1988)}]{meakin1988scaling}%
  \BibitemOpen
  \bibfield  {author} {\bibinfo {author} {\bibfnamefont {P.}~\bibnamefont
  {Meakin}}\ and\ \bibinfo {author} {\bibfnamefont {M.~H.}\ \bibnamefont
  {Ernst}},\ }\href@noop {} {\bibfield  {journal} {\bibinfo  {journal} {Phys.
  Rev. Lett.}\ }\textbf {\bibinfo {volume} {60}},\ \bibinfo {pages} {2503}
  (\bibinfo {year} {1988})}\BibitemShut {NoStop}%
\bibitem [{\citenamefont {Mason}\ \emph {et~al.}(1997)\citenamefont {Mason},
  \citenamefont {Ganesan}, \citenamefont {{Van Zanten}}, \citenamefont
  {Wirtz},\ and\ \citenamefont {Kuo}}]{Mason1997}%
  \BibitemOpen
  \bibfield  {author} {\bibinfo {author} {\bibfnamefont {T.~G.}\ \bibnamefont
  {Mason}}, \bibinfo {author} {\bibfnamefont {K.}~\bibnamefont {Ganesan}},
  \bibinfo {author} {\bibfnamefont {J.~H.}\ \bibnamefont {{Van Zanten}}},
  \bibinfo {author} {\bibfnamefont {D.}~\bibnamefont {Wirtz}},\ and\ \bibinfo
  {author} {\bibfnamefont {S.~C.}\ \bibnamefont {Kuo}},\ }\href@noop {}
  {\bibfield  {journal} {\bibinfo  {journal} {Physical Review Letters}\
  }\textbf {\bibinfo {volume} {79}},\ \bibinfo {pages} {3282} (\bibinfo {year}
  {1997})}\BibitemShut {NoStop}%
\bibitem [{\citenamefont {Zhu}\ \emph {et~al.}(2008)\citenamefont {Zhu},
  \citenamefont {Kundukad},\ and\ \citenamefont {{Van Der Maarel}}}]{Zhu2008}%
  \BibitemOpen
  \bibfield  {author} {\bibinfo {author} {\bibfnamefont {X.}~\bibnamefont
  {Zhu}}, \bibinfo {author} {\bibfnamefont {B.}~\bibnamefont {Kundukad}},\ and\
  \bibinfo {author} {\bibfnamefont {J.~R.}\ \bibnamefont {{Van Der Maarel}}},\
  }\href@noop {} {\bibfield  {journal} {\bibinfo  {journal} {J. Chem. Phys.}\
  }\textbf {\bibinfo {volume} {129}},\ \bibinfo {pages} {1} (\bibinfo {year}
  {2008})}\BibitemShut {NoStop}%
\bibitem [{\citenamefont {Krajina}\ \emph {et~al.}(2017)\citenamefont
  {Krajina}, \citenamefont {Tropini}, \citenamefont {Zhu}, \citenamefont
  {Digiacomo}, \citenamefont {Sonnenburg}, \citenamefont {Heilshorn},\ and\
  \citenamefont {Spakowitz}}]{Krajina2017}%
  \BibitemOpen
  \bibfield  {author} {\bibinfo {author} {\bibfnamefont {B.~A.}\ \bibnamefont
  {Krajina}}, \bibinfo {author} {\bibfnamefont {C.}~\bibnamefont {Tropini}},
  \bibinfo {author} {\bibfnamefont {A.}~\bibnamefont {Zhu}}, \bibinfo {author}
  {\bibfnamefont {P.}~\bibnamefont {Digiacomo}}, \bibinfo {author}
  {\bibfnamefont {J.~L.}\ \bibnamefont {Sonnenburg}}, \bibinfo {author}
  {\bibfnamefont {S.~C.}\ \bibnamefont {Heilshorn}},\ and\ \bibinfo {author}
  {\bibfnamefont {A.~J.}\ \bibnamefont {Spakowitz}},\ }\href@noop {} {\bibfield
   {journal} {\bibinfo  {journal} {ACS Central Science}\ }\textbf {\bibinfo
  {volume} {3}},\ \bibinfo {pages} {1294} (\bibinfo {year} {2017})}\BibitemShut
  {NoStop}%
\bibitem [{\citenamefont {Tanoguchi}\ and\ \citenamefont
  {Murayama}(2018)}]{Tanoguchi2018}%
  \BibitemOpen
  \bibfield  {author} {\bibinfo {author} {\bibfnamefont {M.}~\bibnamefont
  {Tanoguchi}}\ and\ \bibinfo {author} {\bibfnamefont {Y.}~\bibnamefont
  {Murayama}},\ }\href@noop {} {\bibfield  {journal} {\bibinfo  {journal} {AIP
  Advances}\ }\textbf {\bibinfo {volume} {8}} (\bibinfo {year}
  {2018})}\BibitemShut {NoStop}%
\bibitem [{\citenamefont {Smrek}\ \emph {et~al.}(2021)\citenamefont {Smrek},
  \citenamefont {Garamella}, \citenamefont {Robertson-Anderson},\ and\
  \citenamefont {Michieletto}}]{Smrek2021}%
  \BibitemOpen
  \bibfield  {author} {\bibinfo {author} {\bibfnamefont {J.}~\bibnamefont
  {Smrek}}, \bibinfo {author} {\bibfnamefont {J.}~\bibnamefont {Garamella}},
  \bibinfo {author} {\bibfnamefont {R.}~\bibnamefont {Robertson-Anderson}},\
  and\ \bibinfo {author} {\bibfnamefont {D.}~\bibnamefont {Michieletto}},\
  }\href@noop {} {\bibfield  {journal} {\bibinfo  {journal} {Science Advances}\
  }\textbf {\bibinfo {volume} {7}},\ \bibinfo {pages} {1} (\bibinfo {year}
  {2021})}\BibitemShut {NoStop}%
\bibitem [{\citenamefont {Michieletto}\ \emph {et~al.}(2022)\citenamefont
  {Michieletto}, \citenamefont {Neill}, \citenamefont {Weir}, \citenamefont
  {Evans}, \citenamefont {Crist}, \citenamefont {Martinez},\ and\ \citenamefont
  {Robertson-Anderson}}]{Michieletto2022natcomm}%
  \BibitemOpen
  \bibfield  {author} {\bibinfo {author} {\bibfnamefont {D.}~\bibnamefont
  {Michieletto}}, \bibinfo {author} {\bibfnamefont {P.}~\bibnamefont {Neill}},
  \bibinfo {author} {\bibfnamefont {S.}~\bibnamefont {Weir}}, \bibinfo {author}
  {\bibfnamefont {D.}~\bibnamefont {Evans}}, \bibinfo {author} {\bibfnamefont
  {N.}~\bibnamefont {Crist}}, \bibinfo {author} {\bibfnamefont {V.~A.}\
  \bibnamefont {Martinez}},\ and\ \bibinfo {author} {\bibfnamefont {R.~M.}\
  \bibnamefont {Robertson-Anderson}},\ }\href@noop {} {\bibfield  {journal}
  {\bibinfo  {journal} {Nature Communications}\ }\textbf {\bibinfo {volume}
  {13}} (\bibinfo {year} {2022})}\BibitemShut {NoStop}%
\bibitem [{\citenamefont {Fosado}\ \emph {et~al.}(2023)\citenamefont {Fosado},
  \citenamefont {Howard}, \citenamefont {Weir}, \citenamefont {Noy},
  \citenamefont {Leake},\ and\ \citenamefont {Michieletto}}]{Fosado2023}%
  \BibitemOpen
  \bibfield  {author} {\bibinfo {author} {\bibfnamefont {Y.~A.}\ \bibnamefont
  {Fosado}}, \bibinfo {author} {\bibfnamefont {J.}~\bibnamefont {Howard}},
  \bibinfo {author} {\bibfnamefont {S.}~\bibnamefont {Weir}}, \bibinfo {author}
  {\bibfnamefont {A.}~\bibnamefont {Noy}}, \bibinfo {author} {\bibfnamefont
  {M.~C.}\ \bibnamefont {Leake}},\ and\ \bibinfo {author} {\bibfnamefont
  {D.}~\bibnamefont {Michieletto}},\ }\href@noop {} {\bibfield  {journal}
  {\bibinfo  {journal} {Physical Review Letters}\ }\textbf {\bibinfo {volume}
  {130}},\ \bibinfo {pages} {58203} (\bibinfo {year} {2023})}\BibitemShut
  {NoStop}%
\bibitem [{\citenamefont {Roovers}(1988)}]{Roovers1988}%
  \BibitemOpen
  \bibfield  {author} {\bibinfo {author} {\bibfnamefont {J.}~\bibnamefont
  {Roovers}},\ }\href {http://pubs.acs.org/@doi/abs/10.1021/ma00183a049}
  {\bibfield  {journal} {\bibinfo  {journal} {Macromolecules}\ }\textbf
  {\bibinfo {volume} {21}},\ \bibinfo {pages} {1517} (\bibinfo {year}
  {1988})}\BibitemShut {NoStop}%
\bibitem [{\citenamefont {Kapnistos}\ \emph {et~al.}(2008)\citenamefont
  {Kapnistos}, \citenamefont {Lang}, \citenamefont {Vlassopoulos},
  \citenamefont {Pyckhout-Hintzen}, \citenamefont {Richter}, \citenamefont
  {Cho}, \citenamefont {Chang},\ and\ \citenamefont
  {Rubinstein}}]{Kapnistos2008}%
  \BibitemOpen
  \bibfield  {author} {\bibinfo {author} {\bibfnamefont {M.}~\bibnamefont
  {Kapnistos}}, \bibinfo {author} {\bibfnamefont {M.}~\bibnamefont {Lang}},
  \bibinfo {author} {\bibfnamefont {D.}~\bibnamefont {Vlassopoulos}}, \bibinfo
  {author} {\bibfnamefont {W.}~\bibnamefont {Pyckhout-Hintzen}}, \bibinfo
  {author} {\bibfnamefont {D.}~\bibnamefont {Richter}}, \bibinfo {author}
  {\bibfnamefont {D.}~\bibnamefont {Cho}}, \bibinfo {author} {\bibfnamefont
  {T.}~\bibnamefont {Chang}},\ and\ \bibinfo {author} {\bibfnamefont
  {M.}~\bibnamefont {Rubinstein}},\ }\href
  {http://www.ncbi.nlm.nih.gov/pubmed/18953345} {\bibfield  {journal} {\bibinfo
   {journal} {Nature materials}\ }\textbf {\bibinfo {volume} {7}},\ \bibinfo
  {pages} {997} (\bibinfo {year} {2008})}\BibitemShut {NoStop}%
\bibitem [{\citenamefont {Halverson}\ \emph {et~al.}(2012)\citenamefont
  {Halverson}, \citenamefont {Grest}, \citenamefont {Grosberg},\ and\
  \citenamefont {Kremer}}]{Halverson2012}%
  \BibitemOpen
  \bibfield  {author} {\bibinfo {author} {\bibfnamefont {J.~D.}\ \bibnamefont
  {Halverson}}, \bibinfo {author} {\bibfnamefont {G.~S.}\ \bibnamefont
  {Grest}}, \bibinfo {author} {\bibfnamefont {A.~Y.}\ \bibnamefont
  {Grosberg}},\ and\ \bibinfo {author} {\bibfnamefont {K.}~\bibnamefont
  {Kremer}},\ }\href {http://link.aps.org/@doi/10.1103/PhysRevLett.108.038301}
  {\bibfield  {journal} {\bibinfo  {journal} {Phys. Rev. Lett.}\ }\textbf
  {\bibinfo {volume} {108}},\ \bibinfo {pages} {038301} (\bibinfo {year}
  {2012})}\BibitemShut {NoStop}%
\bibitem [{\citenamefont {Zhou}\ \emph {et~al.}(2021)\citenamefont {Zhou},
  \citenamefont {Young}, \citenamefont {Lee}, \citenamefont {Banik},
  \citenamefont {Kong}, \citenamefont {McKenna}, \citenamefont
  {Robertson-Anderson}, \citenamefont {Sing},\ and\ \citenamefont
  {Schroeder}}]{Zhou2021}%
  \BibitemOpen
  \bibfield  {author} {\bibinfo {author} {\bibfnamefont {Y.}~\bibnamefont
  {Zhou}}, \bibinfo {author} {\bibfnamefont {C.~D.}\ \bibnamefont {Young}},
  \bibinfo {author} {\bibfnamefont {M.}~\bibnamefont {Lee}}, \bibinfo {author}
  {\bibfnamefont {S.}~\bibnamefont {Banik}}, \bibinfo {author} {\bibfnamefont
  {D.}~\bibnamefont {Kong}}, \bibinfo {author} {\bibfnamefont {G.~B.}\
  \bibnamefont {McKenna}}, \bibinfo {author} {\bibfnamefont {R.~M.}\
  \bibnamefont {Robertson-Anderson}}, \bibinfo {author} {\bibfnamefont {C.~E.}\
  \bibnamefont {Sing}},\ and\ \bibinfo {author} {\bibfnamefont {C.~M.}\
  \bibnamefont {Schroeder}},\ }\href@noop {} {\bibfield  {journal} {\bibinfo
  {journal} {Journal of Rheology}\ }\textbf {\bibinfo {volume} {65}},\ \bibinfo
  {pages} {729} (\bibinfo {year} {2021})}\BibitemShut {NoStop}%
\bibitem [{\citenamefont {Parisi}\ \emph {et~al.}(2020)\citenamefont {Parisi},
  \citenamefont {Ahn}, \citenamefont {Chang}, \citenamefont {Vlassopoulos},\
  and\ \citenamefont {Rubinstein}}]{Parisi2020}%
  \BibitemOpen
  \bibfield  {author} {\bibinfo {author} {\bibfnamefont {D.}~\bibnamefont
  {Parisi}}, \bibinfo {author} {\bibfnamefont {J.}~\bibnamefont {Ahn}},
  \bibinfo {author} {\bibfnamefont {T.}~\bibnamefont {Chang}}, \bibinfo
  {author} {\bibfnamefont {D.}~\bibnamefont {Vlassopoulos}},\ and\ \bibinfo
  {author} {\bibfnamefont {M.}~\bibnamefont {Rubinstein}},\ }\href@noop {}
  {\bibfield  {journal} {\bibinfo  {journal} {Macromolecules}\ }\textbf
  {\bibinfo {volume} {53}},\ \bibinfo {pages} {1685} (\bibinfo {year}
  {2020})}\BibitemShut {NoStop}%
\bibitem [{\citenamefont {Michieletto}\ \emph {et~al.}(2014)\citenamefont
  {Michieletto}, \citenamefont {Marenduzzo}, \citenamefont {Orlandini},
  \citenamefont {Alexander},\ and\ \citenamefont
  {Turner}}]{Michieletto2014self}%
  \BibitemOpen
  \bibfield  {author} {\bibinfo {author} {\bibfnamefont {D.}~\bibnamefont
  {Michieletto}}, \bibinfo {author} {\bibfnamefont {D.}~\bibnamefont
  {Marenduzzo}}, \bibinfo {author} {\bibfnamefont {E.}~\bibnamefont
  {Orlandini}}, \bibinfo {author} {\bibfnamefont {G.~P.}\ \bibnamefont
  {Alexander}},\ and\ \bibinfo {author} {\bibfnamefont {M.~S.}\ \bibnamefont
  {Turner}},\ }\href {http://dx.@doi.org/10.1039/C4SM00619D} {\bibfield
  {journal} {\bibinfo  {journal} {Soft Matter}\ }\textbf {\bibinfo {volume}
  {10}},\ \bibinfo {pages} {5936} (\bibinfo {year} {2014})}\BibitemShut
  {NoStop}%
\bibitem [{\citenamefont {Soh}\ \emph {et~al.}(2019)\citenamefont {Soh},
  \citenamefont {Klotz}, \citenamefont {Robertson-Anderson},\ and\
  \citenamefont {Doyle}}]{Soh2019a}%
  \BibitemOpen
  \bibfield  {author} {\bibinfo {author} {\bibfnamefont {B.~W.}\ \bibnamefont
  {Soh}}, \bibinfo {author} {\bibfnamefont {A.~R.}\ \bibnamefont {Klotz}},
  \bibinfo {author} {\bibfnamefont {R.~M.}\ \bibnamefont
  {Robertson-Anderson}},\ and\ \bibinfo {author} {\bibfnamefont {P.~S.}\
  \bibnamefont {Doyle}},\ }\href@noop {} {\bibfield  {journal} {\bibinfo
  {journal} {Physical Review Letters}\ }\textbf {\bibinfo {volume} {123}},\
  \bibinfo {pages} {1} (\bibinfo {year} {2019})}\BibitemShut {NoStop}%
\bibitem [{\citenamefont {Annaluru}\ \emph {et~al.}(2014)\citenamefont
  {Annaluru}, \citenamefont {Muller}, \citenamefont {Mitchell}, \citenamefont
  {Ramalingam}, \citenamefont {Stracquadanio}, \citenamefont {Richardson},
  \citenamefont {Dymond}, \citenamefont {Kuang}, \citenamefont {Scheifele},
  \citenamefont {Cooper}, \citenamefont {Cai}, \citenamefont {Zeller},
  \citenamefont {Agmon},\ and\ \citenamefont {Han}}]{Annaluru2014}%
  \BibitemOpen
  \bibfield  {author} {\bibinfo {author} {\bibfnamefont {N.}~\bibnamefont
  {Annaluru}}, \bibinfo {author} {\bibfnamefont {H.}~\bibnamefont {Muller}},
  \bibinfo {author} {\bibfnamefont {L.~A.}\ \bibnamefont {Mitchell}}, \bibinfo
  {author} {\bibfnamefont {S.}~\bibnamefont {Ramalingam}}, \bibinfo {author}
  {\bibfnamefont {G.}~\bibnamefont {Stracquadanio}}, \bibinfo {author}
  {\bibfnamefont {S.~M.}\ \bibnamefont {Richardson}}, \bibinfo {author}
  {\bibfnamefont {J.~S.}\ \bibnamefont {Dymond}}, \bibinfo {author}
  {\bibfnamefont {Z.}~\bibnamefont {Kuang}}, \bibinfo {author} {\bibfnamefont
  {L.~Z.}\ \bibnamefont {Scheifele}}, \bibinfo {author} {\bibfnamefont {E.~M.}\
  \bibnamefont {Cooper}}, \bibinfo {author} {\bibfnamefont {Y.}~\bibnamefont
  {Cai}}, \bibinfo {author} {\bibfnamefont {K.}~\bibnamefont {Zeller}},
  \bibinfo {author} {\bibfnamefont {N.}~\bibnamefont {Agmon}},\ and\ \bibinfo
  {author} {\bibfnamefont {J.~S.}\ \bibnamefont {Han}},\ }\href@noop {}
  {\bibfield  {journal} {\bibinfo  {journal} {Science (New York, N.Y.)}\
  }\textbf {\bibinfo {volume} {344}},\ \bibinfo {pages} {55} (\bibinfo {year}
  {2014})},\ \Eprint {https://arxiv.org/abs/24674868} {arXiv:24674868}
  \BibitemShut {NoStop}%
\bibitem [{\citenamefont {{Del Grosso}}\ \emph {et~al.}(2022)\citenamefont
  {{Del Grosso}}, \citenamefont {Franco}, \citenamefont {Prins},\ and\
  \citenamefont {Ricci}}]{DelGrosso2022}%
  \BibitemOpen
  \bibfield  {author} {\bibinfo {author} {\bibfnamefont {E.}~\bibnamefont {{Del
  Grosso}}}, \bibinfo {author} {\bibfnamefont {E.}~\bibnamefont {Franco}},
  \bibinfo {author} {\bibfnamefont {L.~J.}\ \bibnamefont {Prins}},\ and\
  \bibinfo {author} {\bibfnamefont {F.}~\bibnamefont {Ricci}},\ }\href
  {https://doi.org/10.1038/s41557-022-00957-6} {\bibfield  {journal} {\bibinfo
  {journal} {Nature Chemistry}\ }\textbf {\bibinfo {volume} {14}},\ \bibinfo
  {pages} {600} (\bibinfo {year} {2022})}\BibitemShut {NoStop}%
\bibitem [{\citenamefont {Heinen}\ and\ \citenamefont
  {Walther}(2019)}]{Heinen2019}%
  \BibitemOpen
  \bibfield  {author} {\bibinfo {author} {\bibfnamefont {L.}~\bibnamefont
  {Heinen}}\ and\ \bibinfo {author} {\bibfnamefont {A.}~\bibnamefont
  {Walther}},\ }\href@noop {} {\bibfield  {journal} {\bibinfo  {journal}
  {Science Advances}\ }\textbf {\bibinfo {volume} {5}},\ \bibinfo {pages} {32}
  (\bibinfo {year} {2019})}\BibitemShut {NoStop}%
\end{thebibliography}

\end{document}